\documentclass{aa}
\usepackage{hyperref}
\usepackage{txfonts}
\usepackage{graphicx}
\usepackage[tbtags,fleqn]{amsmath} 
\bibpunct{(}{)}{;}{a}{}{,}

\begin{document}

\title{Signature of the presence of a third body orbiting
  around  XB 1916-053}

\author{R. Iaria\inst{1}, T. Di Salvo\inst{1},  A. F. Gambino\inst{1}, M. Del
Santo\inst{2}, P. Romano\inst{2}, M. Matranga\inst{1},
C. G. Galiano\inst{1}, F. Scarano\inst{3},
A. Riggio\inst{3}, A. Sanna\inst{3}, F. Pintore\inst{3}, L. Burderi\inst{3}}

\institute{Dipartimento di Fisica e Chimica,
Universit\`a di Palermo, via Archirafi 36 - 90123 Palermo, Italy\\
\email{rosario.iaria@unipa.it}
\and 
Istituto Nazionale di Astrofisica, IASF Palermo, Via U. La Malfa 153,
I-90146 Palermo, Italy
\and
Dipartimento di Fisica, Universit\`a degli Studi di Cagliari, SP
Monserrato-Sestu, KM 0.7, Monserrato, 09042 Italy}

\date{\today}

\abstract
{The ultra-compact dipping source \object{XB 1916-053} has an orbital period of
  close to 50 min and a  companion star with a very low mass (less
  than 0.1 M$_{\odot}$).  The orbital period derivative of the source
was estimated to be $1.5(3) \times 10^{-11}$ s/s through analysing the delays
associated with the dip arrival times obtained from observations
spanning 25 years, from 1978 to 2002.} 
{The known orbital period derivative is extremely large and can be
  explained  by invoking an extreme,  non-conservative mass transfer
  rate that is not
  easily justifiable. We extended the analysed data
  from 1978 to 2014, by spanning 37 years, to verify whether  a
  larger sample of data can be fitted with a quadratic term or a 
  different scenario has to be considered. 
}
{ We obtained 27 delays associated with the dip arrival times from
  data covering 37 years and used different models to fit the time
  delays with respect to a constant period model. }
{  We  find that the  quadratic form alone does not fit the data. The
 data are well fitted using   a sinusoidal term plus a quadratic function
 or, alternatively, with a series of sinusoidal terms that can be
 associated  with  a modulation of the dip arrival times due to the
 presence of a third body that has an elliptical orbit.  We infer
that for a conservative mass transfer scenario the modulation of the
delays can be explained by invoking the presence of a third body with
mass between   0.10--0.14 M$_{\odot}$,
orbital period around the X-ray binary system of close to 51 yr
and an eccentricity of $0.28 \pm 0.15$.
In a non-conservative mass transfer scenario we estimate that the 
fraction of matter yielded by the degenerate companion star and
accreted onto the neutron star is $\beta = 0.08$, the neutron star
mass is $\ge 2.2$ M$_{\odot}$,  and the companion star mass is 0.028 M$_{\odot}$.
In this case, we explain the sinusoidal modulation of the delays  by
invoking the presence of a third body with orbital period of 26 yr and 
mass of 0.055 M$_{\odot}$. 
}
{From the analysis of the delays associated with the dip arrival times, 
we find that both in a conservative and non-conservative mass transfer
scenario  we have to invoke the presence of a third
body to explain the observed sinusoidal modulation. 
We propose that XB 1916-053
forms a  hierarchical triple system. 
}
  \authorrunning{R.\ Iaria et al.}

\titlerunning{Is XB 1916-053 part of hierarchical triple system?}

\maketitle

\section{Introduction}
The X-ray source XB 1916-053 is a low-mass X-ray binary (LMXB) showing dips
and  type-I X-ray bursts in its light curves.  Using \textit{OSO 8} data,
\cite{Becker_77} observed type-I X-ray bursts, implying that the
compact source in XB 1916-053 is a neutron star.  Assuming that the
peak luminosity of the X-ray bursts in XB 1916-053 is at the Eddington
limit, \cite{Smale_88} derived a distance to the source of 8.4 kpc or
10.8 kpc, respectively, depending on whether the accreting matter has
cosmic abundances or  is extremely hydrogen-deficient.
\cite{Yoshida_PhD_93} inferred a distance to the source of 9.3 kpc
studying the photospheric radius expansion of the X-ray bursts in XB
1916-053 \citep[see also][]{Barret_96}.  XB 1916-053 was the first
LMXB in which periodic absorption dips
were detected \citep{Walter_82, White_82}.  These dips represent a decrease in
 the count rate in the light curve caused by periodic absorption of the
X-ray emission produced in the inner region of the system. The
photoelectric absorption occurs in a bulge at the outer radius of the
accretion disc where the matter streaming from a companion star
impacts.
 
Accurate analysis of data sets from many X-ray satellites in the last
30 years have found different values for the X-ray period:
\cite{Walter_82} found a period close to 2$\,$985 s, using \textit{Einstein}
data; \cite{White_82} estimated a period of $3\,003.6 \pm 1.8$ s for the
strongest dips, while \cite{Smale_89}, analysing \textit{GINGA} data,
derived a period of $3\,005.0\pm6.6$ s.  \cite{Church_97}, analysing ASCA
data, found an orbital period of $3\,005\pm10$ s.
The X-ray light curve of XB 1916-053 also shows  secondary
dips occurring  approximately half a cycle away from the primary dips
with a certain variability in phase \citep[see][]{Grindlay_89}.
No eclipses were found; this constrains the orbital
inclination of the system  between 60$^\circ$ and  80$^\circ$. 

The optical counterpart of XB1916-053 was discovered by
\cite{Grindlay_87},  a star with a V  magnitude of 21 already noted by
\cite{Walter_82}.  
Using   thermonuclear flash models  of X-ray bursts, \cite{Swank_84} argued 
that the companion star is not hydrogen
exhausted and suggested a companion star mass of 0.1 M$_{\odot}$. 
Furthermore, \cite{Pac} showed that X-ray binary systems with orbital
periods shorter than 81 min cannot contain hydrogen-rich secondary stars.

A modulation in the optical light curve with a period of $3\,027.4 \pm
0.4$ s was discovered by \cite{Grindlay_88}.  The 1\% discrepancy
between the optical and X-ray period of XB 1916-053 was explained by
\cite{Grindlay_88} invoking the presence of a third body with a period
of 2.5 d and a retrograde orbit that influences the matter streaming
from the companion star. The same authors also suggested the
alternative scenario in which the disc bulge precesses around the disc
with a prograde period equivalent to the beat period between the
optical and X-ray period.  \cite{White_89} suggested the possibility
that a precessing elliptical disc exists in XB 1916-053, and that the
variation in the projected area of this disc causes  optical
modulation.  \cite{Callanan_95} showed the stability of the optical
period over seven years.
\begin{figure*}
\centering
\includegraphics[width=8.5cm]{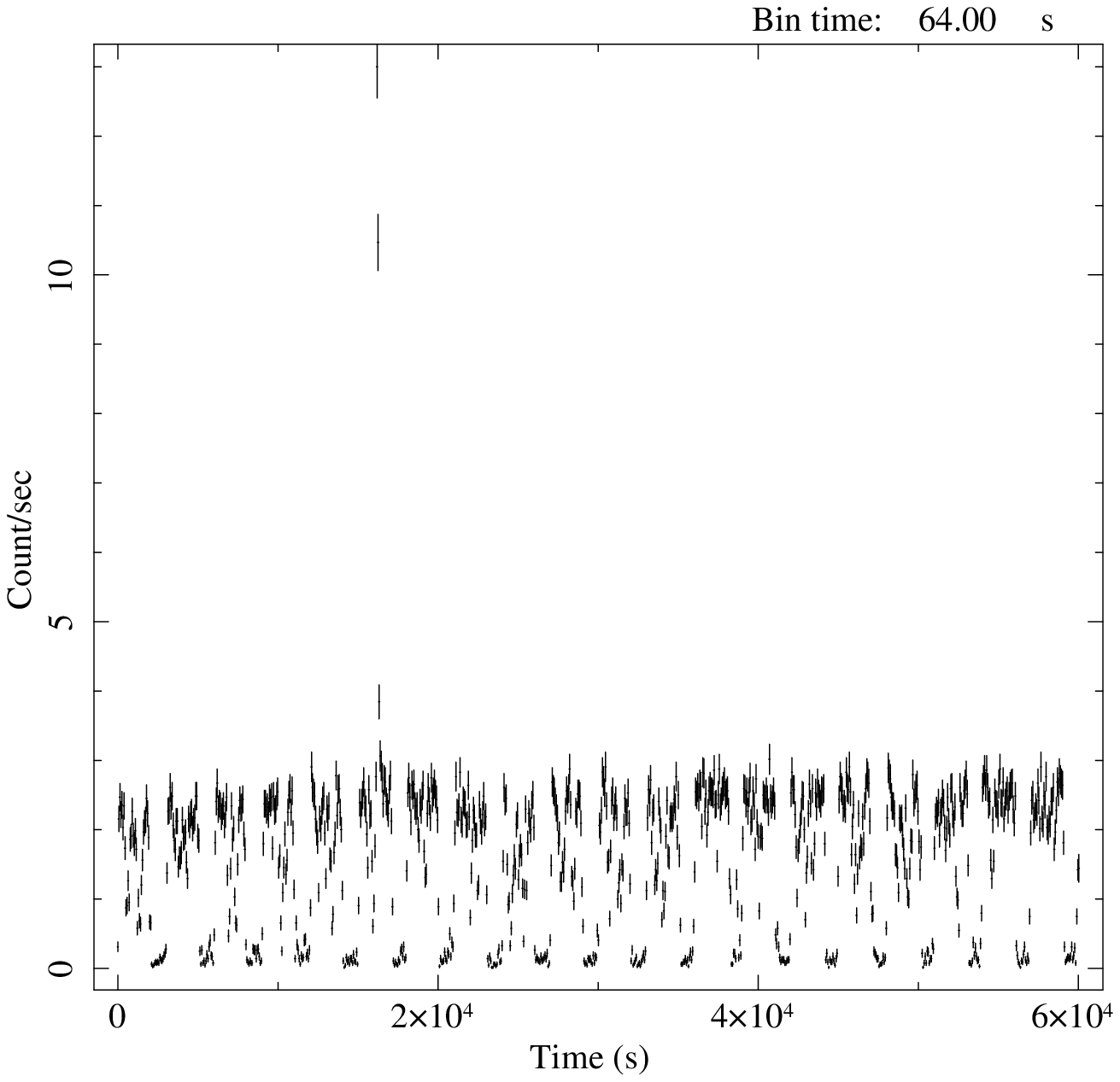}\hspace{0.3truecm}
\includegraphics[width=8.5cm]{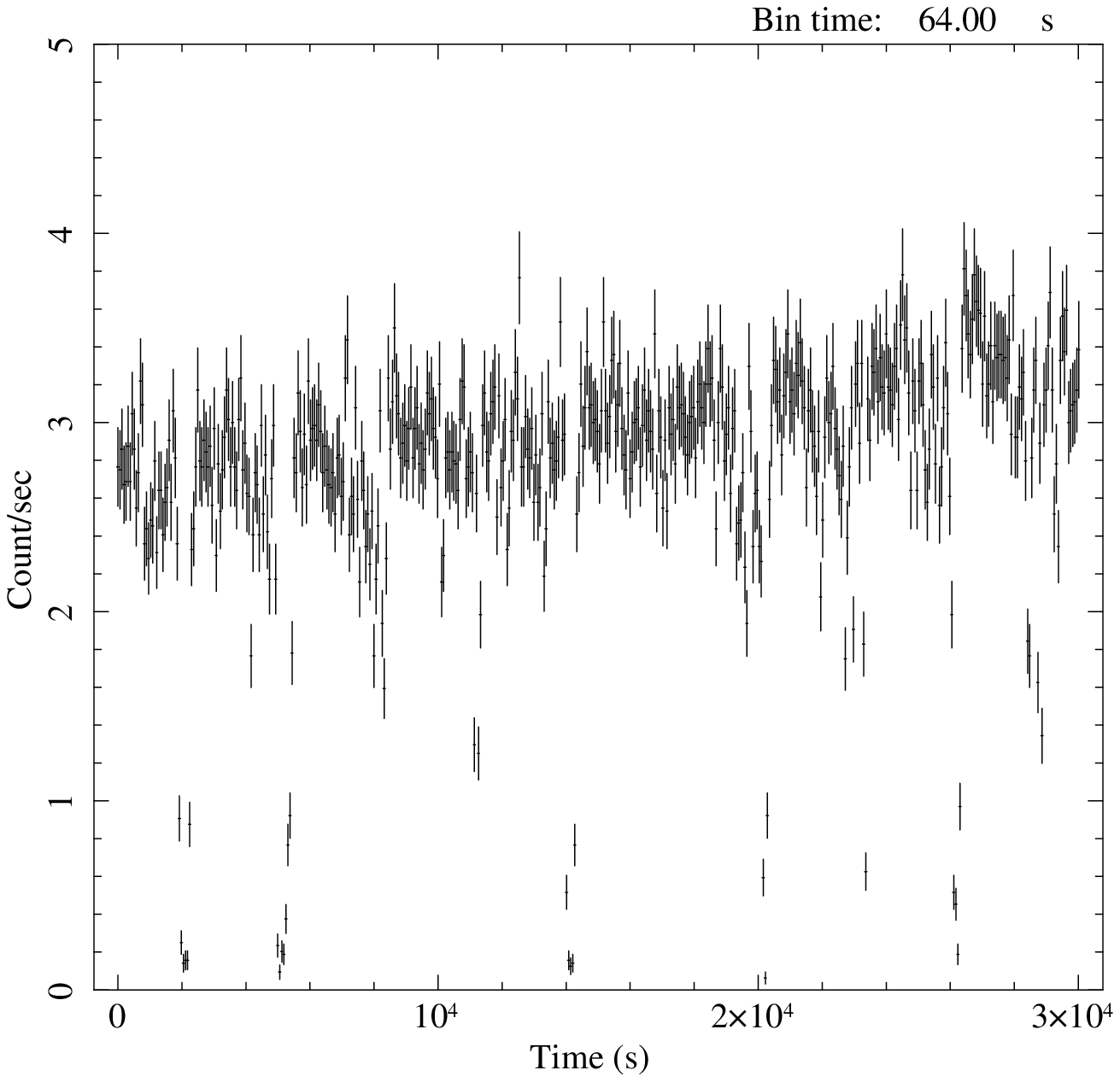}
\caption{Chandra/LEG  light curves of XB 1916-053 during the two
  observations  performed in 2013, i.e.  obsid. 15271 (left) and
  15657 (right). The bin time
  is 64 s. A type-I X-ray burst that  occurred
  during the  obsid. 15271.}
 \label{chandra}
\end{figure*}
\cite{Chou_01}, analysing {\it Rossi X-ray Timing Explorer} (RXTE) data taken in 1996, found several
periodicities including one at $3\,026.23 \pm 3.23$ s, which was similar to the
optical modulation at 3$\,$027 s. The centroid of these peaks in the
periodogram associated with the 3$\,$000 s period implies that there is a
modulation with a fundamental period close to 3.9 d, as already
noticed by \cite{Grindlay_92} also in the optical band. The 
period of 3.9 d is interpreted as the beat period between the
optical and X-ray periods. Furthermore,
\cite{Chou_01}, folding the RXTE light curves at the 3.9 d period,
found changes in the dip shape following this modulation. Those authors
also indicated that the dip-phase change, with a sinusoidal period of
$6.5\pm1.1$ d from Ginga 1990 September observations
\citep{Yoshida_PhD_93,Yoshida_95},  may be  associated with the subharmonic
of the 3.9 d period.  \cite{Retter_02} detected a further
independent X-ray period at 2$\,$979 s in the RXTE light curves of XB
1916-053, which  was mistakenly identified by \cite{Chou_01} with
a 3.9 d sideband of the 3$\,$000 s period. \cite{Retter_02} suggested 
that the  period at 2$\,$979 s  could be explained as a negative super-hump
assuming the   3$\,$000 s period is the orbital period with a
corresponding  beat period of 4.8 d. The same authors suggested that
the  3.9  and 4.8 d periods
could be the apsidal and nodal precession of the accretion disc, respectively.  
 
Finally, the source also showed  a long-term $198.6\pm1.72$
 d periodicity in X-rays 
 \citep{Priedhorsky}, which has not  been confirmed by further
 observations  \citep[see][]{Retter_02}. 
 To date the spin period of the neutron star in  XB
 1916-053 is not known.
\cite{Galloway2}, analysing a Type-I X-ray burst,  discovered a highly
 coherent oscillation drifting from  269.4 Hz up to 272
 Hz. Interpreting the asymptotic frequency of the oscillation in terms
 of a decoupled surface burning layer, the neutron star could have a
 spin period around 3.7 ms.

 \cite{Hu_08} inferred that
 $\dot P_{orb}/P_{orb}=(1.62\pm0.34) \times 10^{-7}$ yr$^{-1}$ by
 analysing archival X-ray data from 1978 to 2002 and adopting a quadratic
 ephemeris to fit the dip arrival times.
\begin{table*}
\scriptsize
  \caption{Observation Log. \label{Tab_obs} }
\begin{center}
\begin{tabular}{l l l l l l}          
\hline\hline 
Point & Satellite/Instrument  &               
Observation  &               
Start Time &
Stop Time  &
$T_{fold}$ \\

& & & (UT) & (UT) & (MJD,TDB)\\
\hline                                             

1 & OSO-8/GCXSE &
&
1978 Apr 07 21:16:05&
1978 Apr 14 22:20:37&
43$\,$609.408575724435
\\

2 & Einstein/IPC &
&
1979 Oct 22 04:52:01&
1979 Oct 22 06:58:30&
44$\,$168.24670380917  \\

3 & Einstein/IPC &
&
1980 Oct 11 04:08:51&
1980 Oct 11 09:07:19&
44$\,$523.27644368849 \\

4 & EXOSAT/ME &
 & 
1983 Sep 17 15:07:25&
1983 Sep 17 21:29:49 &
45$\,$594.765324269885\\

5 & EXOSAT/ME &
 & 
1985 May 24 12:26:21&
1985 May 24 21:30:23&
46$\,$209.612747685185
\\

6 & EXOSAT/ME &
 & 
1985 Oct 13 13:53:16&
1985 Oct 13 22:34:04 &
46$\,$351.75944524423
\\

7 & Ginga/LAC &
&
1988 Sep 09 15:47:56&
1988 Sep 10 16:01:16&
47$\,$414.165911835925\\

8 & Ginga/LAC &
&
1990 Sep 11 15:04:35&
1990 Sep 13 09:18:11&
48$\,$146.51075733274\\

9 & ROSAT/PSPC &
RP400274N00&
1992 Oct 17 13:05:47&
1992 Oct 19 15:24:20&
48$\,$913.59379352164\\

10 & ASCA/GIS3 &
40004000&
1993 May 02 18:11:00&
1993 May 03 09:46:17&
49$\,$110.082393510115\\

11 & RXTE/PCA&
P10109-01-01-00, P10109-01-02-00, &
1996 Feb 02 00:14:56&
1996 May 23 11:20:00&
50$\,$174.74129123185\\
 & & 
P10109-01-04-01, P10109-01-04-00, & & \\
 & & 
P10109-02-01-00, P10109-02-02-00, & & \\
 & & 
P10109-02-03-00, P10109-02-04-00, & & \\
 & & 
P10109-02-05-00, P10109-02-06-00, & & \\
 & & 
P10109-02-07-00,P10109-02-08-00,  & & \\
 & & 
P10109-02-09-00, P10109-02-10-00,  & & \\
 & & 
P10109-02-10-02  & & \\

12 & RXTE/PCA&
P10109-01-05-00, P10109-01-06-00, &
1996 Jun 01 17:38:40 &
1996 Oct 29 11:00:34 &
50$\,$310.596956288645\\
 & & 
P10109-01-07-00, P10109-01-08-00, & & \\
 & & 
P10109-01-09-00 & & \\

13& BeppoSAX/MECS&
20106001 &
1997 Apr 27 21:00:06 &
1997 Apr 28 19:51:02&
50$\,$566.35264963594\\

14& RXTE/PCA&
P30066-01-01-04, P30066-01-01-00,  &
1998 Jun 23 23:06:40&
1998 Jul 20 15:35:55&
51$\,$001.306447481845\\
 & & 
P30066-01-01-01, P30066-01-01-02,  & & \\
 & & 
P30066-01-01-03, P30066-01-02-00,  & & \\
 & & 
P30066-01-02-01, P30066-01-02-02, & & \\
 & & 
P30066-01-02-03 & & \\

15& RXTE/PCA&
P30066-01-02-04, P30066-01-02-07,  &
1998 Jul 21 07:11:44&
1998 Sep 16 02:52:32&
51$\,$043.70980975036\\
 & & 
P30066-01-02-08, P30066-01-03-00, & & \\
 & & 
P30066-01-03-01, P30066-01-03-02, & & \\
 & & 
P30066-01-03-03, P30066-01-03-04, & & \\
 & & 
P30066-01-03-05, P30066-01-04-00 & & \\

16& RXTE/PCA&
P30066-01-05-01, P30066-01-05-00,  &
2001 May 27 08:14:47&
2001 Jul 01 19:15:33&
52$\,$074.07302734295 \\
 & & 
P30066-01-06-00, P30066-01-06-01,   & & \\
 & & 
P30066-01-07-00, P30066-01-07-01   & & \\

17& BeppoSAX/MECS&
21373002 &
2001 Oct 01 03:40:16 &
2001 Oct 02 07:01:06&
52$\,$183.72270184033\\

18& RXTE/PCA&
P50026-03-01-00, P50026-03-01-01 &
2001 Oct 01 10:35:44 &
2001 Oct 01 22:16:03&
52$\,$183.684644754605\\

19& RXTE/PCA&
P70034-02-02-01, P70034-02-02-00 &
2002 Sep 25 00:43:12 &
2002 Sep 25 09:31:12&
52$\,$542.21332826887\\

20& XMM/Epic-pn  &
0085290301 &
2002 Sep 25 04:18:29&
2002 Sep 25 08:28:27&
52$\,$542.266295747205\\

21& INTEGRAL/JEM-X  &
 &
2003 Nov 09 09:04:11&
2003 Nov 20 12:18:01&
52$\,$957.945226848465\\

22& Chandra/HETGS &
4584&
2004 Aug 07 02:34:45&
2004 Aug 07 16:14:53&
53$\,$224.59478392645\\

23& Suzaku/XIS0 &
401095010 &
2006 Nov 08 06:09:51&
2006 Nov 09 02:42:02&
54$\,$048.3655207864\\

24& RXTE/PCA&
P95093-01-01-00, P95093-01-01-01 &
2010 Jun 19 13:41:52&
2010 Jun 21 07:21:46 &
55$\,$367.43875650959\\

25& Chandra/LETGS&
15271, 15657 &
2013 Jun 15 13:56:17&
2013 Jun 18 05:13:17&
56$\,$459.89915961875 \\

26& Swift/XRT  &
00033336001  &
2014 Jul 15 08:04:57&
2014 Jul 15 22:36:46&
56$\,$853.63959388178\\

27 & Suzaku/XIS0& 
409032010,  409032020 &
2014 Oct 14 16:49:56     &
2014 Oct 22 2:40:56        &
56$\,$949.56345974802\\ 
\hline                                             
\end{tabular}
\end{center}

\end{table*}
In this work, we update the previously determined ephemeris using data 
from 1978 to 2014. We show that the quadratic ephemeris does not 
fit the dip arrival times and find that a sinusoidal component is
necessary to fit the delays. We suggest  the presence of a third body
that influences the orbit of the X-ray binary system XB 1916-053.

\section{Observations and data reduction}
We used all the available X-ray archival data of XB 1916-053 to study the
long-term change of its orbital period. The last ephemeris of the
source was reported by \cite{Hu_08} who used archival data from 1978
to 2002. We analysed more than 37 years of observational data from
1978 to 2014.  The data have been obtained from the
HEASARC (NASA’s High Energy Astrophysics Science Archive Research
Center) website and have been reduced using the standard
procedures. In particular, we reanalysed the data used by \cite{Hu_08},
collected from 1998 to 2002, and added new data spanning up to 2014
(see Tab. \ref{Tab_obs}). We
obtained 27 points from all the analysed observations.  The data
collected by RXTE, Ginga, EXOSAT, Einstein, and OSO-8 were downloaded
from HEASARC in light-curve format.  We used the standard-1 RXTE/PCA
background-subtracted light curves, which include all the energy
channels and have a time resolution of 0.125 s. All the pointing
observations were used except for P70034-02-01-01, P70034-02-01-00, and
P93447-01-01-00 due to the absence of  dips  in the corresponding light
curves.  The EXOSAT/ME light curves cover the energy range
between 1 and 8 keV and have a time bin of 16 s. The 
Ginga/LAC light curves cover the 2-17 keV energy band. We only used 
the data from the top layer and the light curves binned at 16 s.  We
downloaded the ROSAT/PSPC events, and extracted the corresponding
light curve using the FTOOLS {\tt xselect}.  The Medium Energy
Concentrator Spectrometer (MECS) onboard the BeppoSAX satellite
observed XB 1916-053 two times, in 1997 Apr 27-28 and 2001 Oct
01-02. Using  {\tt xselect}, we extracted the source light
curves from a circular region centred on the
source and with a radius of 4$\arcmin$, no energy filter was applied to
the data. The BeppoSAX/MECS light curves were obtained using a bin time
of 2 sec. ASCA observed XB 1916-053 in 1993 May 02-03; we used the
events collected by the GIS3 working in medium bit rate to extract the
corresponding light curve.
The OSO-8 light curve was obtained using the combined observation of
the B and C detectors of  the GSFC Cosmic X-ray Spectroscopy
experiment (GCXSE). 
The light curve covers the 2-60 keV energy
range. The Einstein light curve was obtained  from events collected by 
the  Image Proportional Counter (IPC) in the 0.2-3.5 keV energy
range.

\begin{table*}
\scriptsize
  \caption{Best-fit parameters obtained fitting the  dips in the folded light curves. \label{Tab_par} }
\begin{center}
\begin{tabular}{l l c c  c c c c c c}          
\hline\hline 
Point  &               
Phase Interval  &               
$C_1$  &
$C_2$ &
$C_3$&
$\phi_1$&
$\phi_2$&
$\phi_3$&
$\phi_4$&
$\chi^2_{red}(d.o.f.)$ \\

       &
       &
count s$^{-1}$  &
count s$^{-1}$  &
 count s$^{-1}$ &
     &
     &
     &
     &
     \\


\hline             
1&
0.7-1.7 &
$7.23 \pm 0.06$   &
$6.01 \pm 0.09$   &
$7.59 \pm 0.06$   &
$1.086 \pm 0.012$  &
$1.113 \pm 0.012$  &
$1.290 \pm 0.016$  &
$1.388 \pm 0.015$  &
1.64(193)\\
 
2&
0.7-1.7 &
$5.74 \pm 0.06$   &
$2.94 \pm 0.13$   &
$5.75 \pm 0.11$   &
$1.133 \pm 0.004$  &
$1.170 \pm 0.004$  &
$1.217 \pm 0.005$  &
$1.263 \pm 0.006$  &
1.44(152)\\

3&
0.1-1 &
$11.24 \pm 0.08$   &
$8.48 \pm 0.12$   &
$11.14 \pm 0.09$   &
$0.416 \pm 0.006$   &
$0.450 \pm 0.005$  &
$0.574 \pm 0.006$  &
$0.603 \pm 0.008$  &
1.90(194)\\

4&
0.8-1.8 &
$23.02 \pm 0.12$   &
$13.7 \pm 0.4$   &
$23.73 \pm 0.11$   &
$1.203 \pm 0.003$  &
$1.239 \pm 0.004$  &
$1.277 \pm 0.003$  &
$1.321 \pm 0.003$  &
1.91(194)\\

5&
0.095-0.8 &
$14.6 \pm 0.4$   &
$8.9 \pm 0.3$   &
$17.7 \pm 0.3$   &
$0.251 \pm 0.015$  &
$0.329 \pm 0.015$  &
$0.500 \pm 0.008$  &
$0.576 \pm 0.008$  &
4.70(114)\\

6&
0.1-0.8&
$25.6\pm 0.2$   &
$20.3 \pm 0.4$   &
$27.1 \pm 0.2$   &
$0.454 \pm 0.007$  &
$0.487 \pm 0.008$  &
$0.561 \pm 0.007$  &
$0.605 \pm 0.007$  &
3.03(133)\\

7&
0.5-1.1 &
$72.9 \pm 1.1$   &
$42.8 \pm 1.0$   &
$75.6 \pm 1.0$   &
$0.658 \pm 0.004$  &
$0.680 \pm 0.004$  &
$0.840 \pm 0.009$  &
$0.940 \pm 0.010$  &
16.9(84)\\

8&
0.5-1.2 &
$107.1 \pm 1.0$   &
$60.6 \pm 1.3$   &
$106.3 \pm 0.8$   &
$0.657 \pm 0.007$  &
$0.805 \pm 0.008$  &
$0.903 \pm 0.004$  &
$0.954 \pm 0.005$  &
47.8(138)\\

9&
0-1 &
$6.05 \pm 0.07$   &
$1.60 \pm 0.06$   &
$6.15 \pm 0.07$   &
$0.340 \pm 0.008$  &
$0.535 \pm 0.006$  &
$0.640 \pm 0.004$  &
$0.752 \pm 0.005$  &
4.77(294)\\

10&
0.6-1.4 &
$9.3 \pm 0.3$   &
$0.85 \pm 0.05$   &
$9.3 \pm 0.3$   &
$0.765 \pm 0.004$  &
$0.809 \pm 0.002$  &
$1.022 \pm 0.003$  &
$1.100 \pm 0.006$  &
12.08(234)\\

11&
0-1 &
$41.00 \pm 0.11$   &
$31.35 \pm 0.11$   &
$45.06 \pm 0.08$   &
$0.207 \pm 0.004$  &
$0.323 \pm 0.004$  &
$0.499 \pm 0.002$  &
$0.611 \pm 0.003$  &
12.61(506)\\

12&
0.1-1.1 &
$36.60 \pm 0.10$   &
$21.69 \pm 0.13$   &
$37.45 \pm 0.11$   &
$0.443 \pm 0.003$  &
$0.569 \pm 0.003$  &
$0.710 \pm 0.002$  &
$0.808 \pm 0.002$  &
11.33(505)\\

13&
0.95-1.95 &
$0.986 \pm 0.013$   &
$0.035 \pm 0.003$   &
$0.982 \pm 0.014$   &
$1.238 \pm 0.004$  &
$1.340 \pm 0.002$  &
$1.537 \pm 0.002$  &
$1.647 \pm 0.005$  &
2.14(249)\\

14&
0.24-0.75 &
$27.6 \pm 0.2$   &
$14.55 \pm 0.10$   &
$27.7 \pm 0.2$   &
$0.313 \pm 0.003$  &
$0.433 \pm 0.002$  &
$0.582 \pm 0.002$  &
$0.705 \pm 0.003$  &
11.10(255)\\

15&
0.15-1 &
$36.51 \pm 0.10$   &
$25.3 \pm 0.2$   &
$38.31 \pm 0.08$   &
$0.381 \pm 0.004$  &
$0.566 \pm 0.005$  &
$0.598 \pm 0.003$  &
$0.738 \pm 0.003$  &
10.43(420)\\

16&
0.35-0.9 &
$25.4 \pm 0.2$   &
$17.52 \pm 0.10$   &
$26.9 \pm 0.2$   &
$0.419 \pm 0.004$  &
$0.489 \pm 0.004$  &
$0.697 \pm 0.003$  &
$0.763 \pm 0.003$  &
1.90(274)\\

17&
0.9-1.9 &
$1.030 \pm 0.009$   &
$0.31 \pm 0.02$   &
$1.016 \pm 0.007$   &
$1.045 \pm 0.004$  &
$1.160 \pm 0.004$  &
$1.178 \pm 0.004$  &
$1.272 \pm 0.004$  &
1.11(249)\\

18&
0-1 &
$23.75 \pm 0.14$   &
$11.4 \pm 0.3$   &
$24.40 \pm 0.13$   &
$1.092 \pm 0.002$  &
$1.113 \pm 0.002$  &
$1.168 \pm 0.005$  &
$1.281 \pm 0.005$  &
1.57(505)\\

19&
0.6-1.6 &
$28.8 \pm 0.2$   &
$20.1 \pm 0.2$   &
$29.8 \pm 0.2$   &
$0.953 \pm 0.004$  &
$0.985 \pm 0.004$  &
$1.171 \pm 0.006$  &
$1.249 \pm 0.006$  &
3.06(505)\\

20&
0.1-1.1 &
$69.9 \pm 0.5$   &
$30.4 \pm 0.5$   &
$72.4 \pm 0.5$   &
$0.406 \pm 0.004$  &
$0.506 \pm 0.003$  &
$0.671 \pm 0.003$  &
$0.730 \pm 0.003$  &
17.84(505)\\

21&
0.2-1.2 &
$ 0.0284 \pm 0.0002$   &
$0.0256 \pm 0.0004$   &
$0.0284 \pm  0.0002$   &
$0.57 \pm 0.02$  &
$0.63 \pm 0.02$  &
$0.708 \pm 0.014$  &
$0.733 \pm 0.011$  &
0.766(144)\\

22&
0.3-1.3 &
$ 9.92 \pm 0.03$   &
$8.5 \pm 0.2$   &
$10.04 \pm  0.03$   &
$0.796 \pm 0.005$  &
$0.835 \pm 0.007$  &
$0.844 \pm 0.013$  &
$0.923 \pm 0.007$  &
1.71(144)\\

23 &
0.9-1.9 &
$ 15.10 \pm 0.05$   &
$11.4 $ (fixed)   &
$15.61 \pm  0.05$   &
$1.263 \pm 0.006$  &
$1.38 $ (fixed)  &
$1.38 $ (fixed)  &
$1.521 \pm 0.006$  &
2.41(505)\\

24&
0.9-1.9 &
$ 30.18 \pm 0.13$   &
$23.1 \pm 0.3$   &
$20.49 \pm  0.15$   &
$1.322 \pm 0.003$  &
$1.344 \pm 0.003$  &
$1.499 \pm 0.002$  &
$1.451 \pm 0.002$  &
1.22(505)\\

25&
0.09-0.65 &
$ 2.52 \pm 0.03$   &
$0.759 \pm 0.010$   &
$2.69 \pm  0.04$   &
$0.194 \pm 0.003$  &
$0.314 \pm 0.002$  &
$0.525 \pm 0.002$  &
$0.598 \pm 0.003$  &
2.39(567)\\

26&
0.6-1.6 &
$ 12.61 \pm 0.11$   &
$4.4 \pm 0.6$   &
$12.5 \pm  0.2$   &
$1.025 \pm 0.010$  &
$1.27 \pm 0.02$  &
$1.295 \pm 0.004$  &
$1.307 \pm 0.004$  &
4.08(171)\\

27&
0.84-1.6 &
$ 5.08 \pm 0.04$   &
$1.607 \pm 0.015$   &
$5.18 \pm  0.02$   &
$0.909 \pm 0.003$  &
$1.089 \pm 0.002$  &
$1.2080 \pm 0.0011$  &
$1.2904 \pm 0.0014$  &
3.29(382)\\

\hline                                             
\end{tabular}
\end{center}
\end{table*}
We applied  barycentre corrections to the whole data set  adopting  the source 
position of XB 1916-053 shown by \cite{Iaria_06}.  
 For the RXTE/PCA  light curves we used  
the ftools {\tt faxbary}. 
The barycentre  corrections for the ASCA and ROSAT  data were obtained using the ftool 
{\tt timeconv} and  the tool  {\tt bct+abc}, respectively.
All the other data sets were corrected using the ftool {\tt
  earth2sun}. Finally, 
we excluded the time intervals containing X-ray bursts from each  analysed 
light {\bf curve}. 
\begin{figure}
\resizebox{\hsize}{!}{\includegraphics{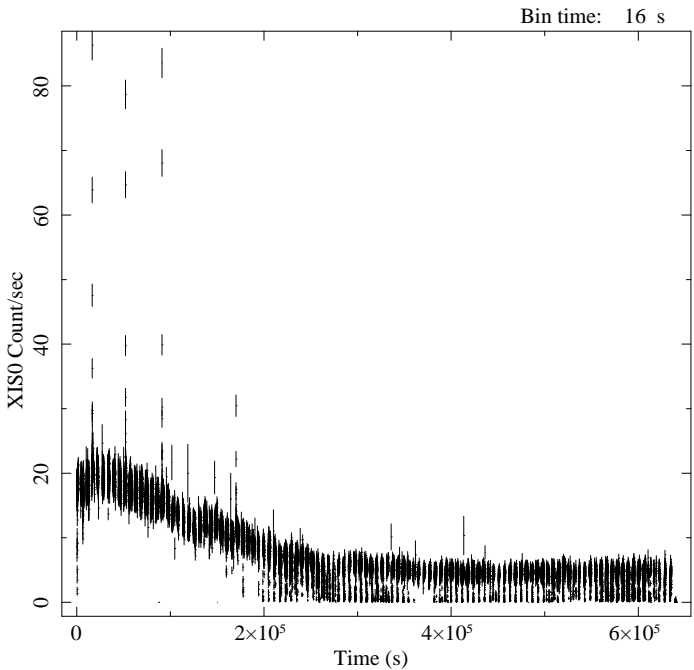}}
\caption{Suzaku/XIS0 light curve of XB 1916-053 during the long
  observation  on 2014 Oct.}
 \label{lightsuz}
\end{figure}

The Chandra satellite observed XB 1916-053 three times. The first time
was on 2004 Aug 07 from 2:34:45 to 16:14:53 UT (obsid 4584). The
observation had a total integration time of 50 ks and was performed in
timed graded mode. The spectroscopic analysis of this data set was
discussed by \cite{Iaria_06}. We reprocessed the data and applied the
barycentre corrections to the event-2 file using the Chandra
Interactive Analysis of Observations (CIAO) tool {\tt axbary}. In
addition,  we
extracted the summed first-order medium energy grating (MEG) and high
energy grating (HEG) light curves filtered in the 0.5-10 keV energy range using
the CIAO tool {\tt dmextract}.
The last two Chandra observations of  XB 1916-053 
(obsid 15271 and 15657)   were performed  between 2013  June 15 13:56
and  June 18 5:13 UT and have  exposure times of 60 and 30 ks,
respectively.  We reprocessed the data and  applied the
barycentre corrections to the event-2 file using  {\tt
  axbary}. Moreover, we  
extracted the first-order  low energy grating (LEG)  light curve in
the 0.5-5 keV energy range using
{\tt dmextract}. We show the Chandra/LEG light curve in 
Fig. \ref{chandra}.  Very intense dipping activity is present during
the two observations.  A  type-I  
burst occurred  during the obsid. 15271.

 \begin{table}
  \caption{Journal of the X-ray dip arrival times of XB 1916-053. \label{Tab1} }
\begin{center}
\begin{tabular}{l l  r  r }          
\hline\hline 
Point &
Dip Time  &               
Cycle  &
Delay  \\
 
  &               
 (MJD;TDB) &               
  &
(s)\\

\hline                        
1 &
43$\,$609.4168(12) &
-187$\,$551 &
$772 \pm 74$\\

2 &
44$\,$168.2535(5)  &
-171$\,$460 &
$792 \pm 28$\\

3 &
44$\,$523.2941(5)   &
-161$\,$237 &
$641 \pm 42$\\

4 &
45$\,$594.7744(3)  &
-130$\,$385  &
$449 \pm 18$\\

5 &
46$\,$209.6271(13) &
-112$\,$681  &
$193 \pm 112$\\

6 &
46$\,$351.7778(9)   &
-108$\,$588 &
$352 \pm 52$\\

7 &
47$\,$414.193(2)&
-77$\,$997  &
$162 \pm 132$\\

8 &
48$\,$146.539(3)&
-56$\,$910&
 $47 \pm 182$\\

9 &
48$\,$913.6127(10)   &
-34$\,$823&
$-140 \pm 59$\\

10 &
49$\,$109.1148(12)  &
-29$\,$165&
$-48 \pm 76$\\

11 &
50$\,$174.7555(5)  &
1$\,$490&
$-50 \pm 46$ \\

12 &
50$\,$310.6187(4) &
 5$\,$402 &
 $-17 \pm 37$ \\

13 &
50$\,$566.3680(4)  &
12$\,$766&
$-69 \pm 39$ \\

14 &
51$\,$001.3241(5)  &
25$\,$290&
$ -15 \pm 40$ \\

15 &
51$\,$043.7292(5)  &
26$\,$511&
$ -9 \pm 45$ \\

16 &
52$\,$074.0935(3)  &
 56$\,$179&
$151 \pm 29$ \\

17 &
52$\,$183.7349(3)  &
 59$\,$336 &
$107 \pm 28$ \\

18 &
52$\,$183.7008(2)  &
 59$\,$335 &
$162 \pm 19$ \\

19 &
 52$\,$542.2168(4)  &
 69$\,$658 &
 $227 \pm 39$ \\

20 &
 52$\,$542.2860(11) &
  69$\,$660 &
 $202 \pm 98$ \\

21 &
52$\,$957.9679(8)  &
  81$\,$629 &
 $327 \pm 69$ \\

22 &
53$\,$224.6246(4)  &
 89$\,$307  &
 $467 \pm 34$ \\

23 &
 54$\,$048.3791(5)  &
 113$\,$026  &
 $411 \pm 39$ \\

24 &
55$\,$367.45218(15)   &
  151$\,$007  &
 $593 \pm 13$ \\

25 &
56$\,$459.9129(3)   &
 182$\,$463   &
 $ 721 \pm 20$ \\

26 &
56$\,$853.6454(8)  &
 193$\,$800  &
 $821 \pm 67 $ \\

27 &
56$\,$949.84670(10)   &
 196$\,$570  &
 $814 \pm 8 $ \\

\hline                                             
\end{tabular}
\end{center}

{\small \sc Note} \footnotesize--- Epoch of reference $50\,123.00873$ MJD, 
orbital period $3\,000.6511$ s.
\end{table}

The X-ray Multi-Mirror Mission-Newton (XMM-Newton) observed XB
1916-053 on 2002 Sep 25 from 3:55 to 8:31 UT and the European Photon
Imaging Camera (EPIC-pn)  collected data, in timing mode, over $\sim 17$ ks
of exposure.  An extensive study of this
observation was performed by \cite{Boirin_xmm}.  We reprocessed the
data, extracted the 0.5-10 keV light curve,  and
applied barycentre  corrections to the times of the EPIC-pn events with
the Science Analysis Software (SAS) tool {\tt barycen}.  

Suzaku observed  XB 1916-053 twice, the first time on 2006 Nov 8
(obsid. 401095010) and the second time from 2014 Oct 14 to 22
(obsid. 409032010 and 409032020). The first observation has already
been analysed by \cite{zhang_2014}, while a study of the second observation
has not been published yet.  For both  observations, we extracted the
 X-ray Imaging Spectrometer 0 (XIS0) events from a circular region centred on the source and with
a radius of 130\arcsec.  We applied the
barycentre corrections to the events with the Suzaku tool {\tt
  aebarycen}. We do not show the light curve of the first Suzaku
observation since it was  already shown by   \cite{zhang_2014} (Fig. 1 in
their paper), however, we show in Fig. \ref{lightsuz} the XIS0 light
curve of the observation performed in 2014 Oct. 
The light curve indicates that a bursting activity is present in the
first 200 ks of the observation and the persistent count rate
decreases from 20  to 10 counts s$^{-1}$.  In the second part of the
observation, the persistent count rate is quite constant at 7 counts s$^{-1}$ and
an intense dipping activity is present. For the aim of this work, we
selected and used  the events from 250 ks to the end of the observation.

{\it Swift}/XRT data were obtained as target of opportunity
observations  performed on 2014 Jul 15
from 07:55:53 to 22:27:58 UT (ObsID 00033336001) for a total on-source
exposure of $\sim6.3$ ks and on 2014 Jul 21 from 07:32:00 to 16:11:5 UT
(ObsID 00033336002) for a total on-source exposure of $\sim9.0$ ks.
The count rate in the first observation reaches 15 counts s$^{-1}$,
with a mean at about 10 counts s$^{-1}$, due to the dips seen down to
2 counts s$^{-1}$; the second observation shows no dips and has a mean
count rate of 7 counts s$^{-1}$. Since the data from ObsID 00033336002
do not show dips we only used  the first observation in our analysis.
The XRT data were processed with standard procedures ({\tt
  xrtpipeline} v0.13.1), and with standard filtering and screening criteria with
{\tt FTOOLS} (v6.16).  Source events (selected in grades 0--2) were
accumulated within a circular region with a radius of 20 pixels (1
pixel $\sim 2.36$\arcsec).  
For our timing analysis, we also
converted the event arrival times to the solar system barycentre with
{\tt barycorr}.

We selected a public data set of {\it INTErnational Gamma-Ray
  Astrophysics Laboratory} \citep[INTEGRAL][]{winkler_03}
observations performed in staring mode on XB 1916-053.  Then, we analysed the data
collected by the X-ray telescope JEM-X2 \citep{lund_03}.  A total amount of 87  pointings (the total observation
elapsed time is $\sim 310$ ks) covered the INTEGRAL revolutions 131,
133, and 134, which were carried out on 2003 November 9-20.  We performed the JEM-X2
data analysis using standard procedures within
the Offline Science Analysis software (OSA10.0) distributed by the
ISDC \citep{cour_03}.  We extracted the light curves with a 16 seconds bin-size
 in the energy range 3-10 keV, and after that we applied
the barycentre corrections to the events using
the tool {\it \textup{\textup{\textit{barycent}}}}.

\section{Data analysis}
We analysed 27 light curves and folded the barycentric-corrected light
curves using a trial time of reference and orbital period, $T_{fold}$
and $P_0$, respectively.  For each light curve, the value of
$T_{fold}$ is defined as the average value between the corresponding
start and stop time.  We fitted the dips 
with a simple model consisting of a step-and-ramp function, where the
count rates before, during, and after the dip are constant and the
intensity changes linearly during the dip transitions. This model
involves seven parameters: the count rate before, during, and after the
dip, called $C_1$, $C_2$, and $C_3$, respectively; the phases of the
start and stop time of the ingress ($\phi_1$ and $\phi_2$), and,
finally, the phases of the start and stop time of the egress ($\phi_3$
and $\phi_4$).
The phase corresponding to the dip arrival time $\phi_{dip}$ is
estimated as $\phi_{dip} = (\phi_4+\phi_1)/2$.  The corresponding dip
arrival time is given by $t_{dip} = T_{fold} + \phi_{dip} P_0 $. To be
more conservative, we scaled the error associated with $\phi_{dip}$ by
the factor $\sqrt{\chi^2_{red}}$ to take  a value of
 $\chi^2_{red}$
of the best-fit model larger than 1 into account.
 To obtain the delays with respect to a constant period reference, we used the
values of the period $P_0=3\,000.6511$ s and reference epoch
$T_0=50\,123.00873$ MJD reported in \cite{Hu_08}.
We show the values of $T_{fold}$ in Tab. \ref{Tab_obs}. The best-fit 
parameters of the step-and-ramp function  and the
corresponding $\chi^2_{red}$  are shown in Tab.  \ref{Tab_par}.
\begin{figure*}
\centering
\includegraphics[width=8.5cm]{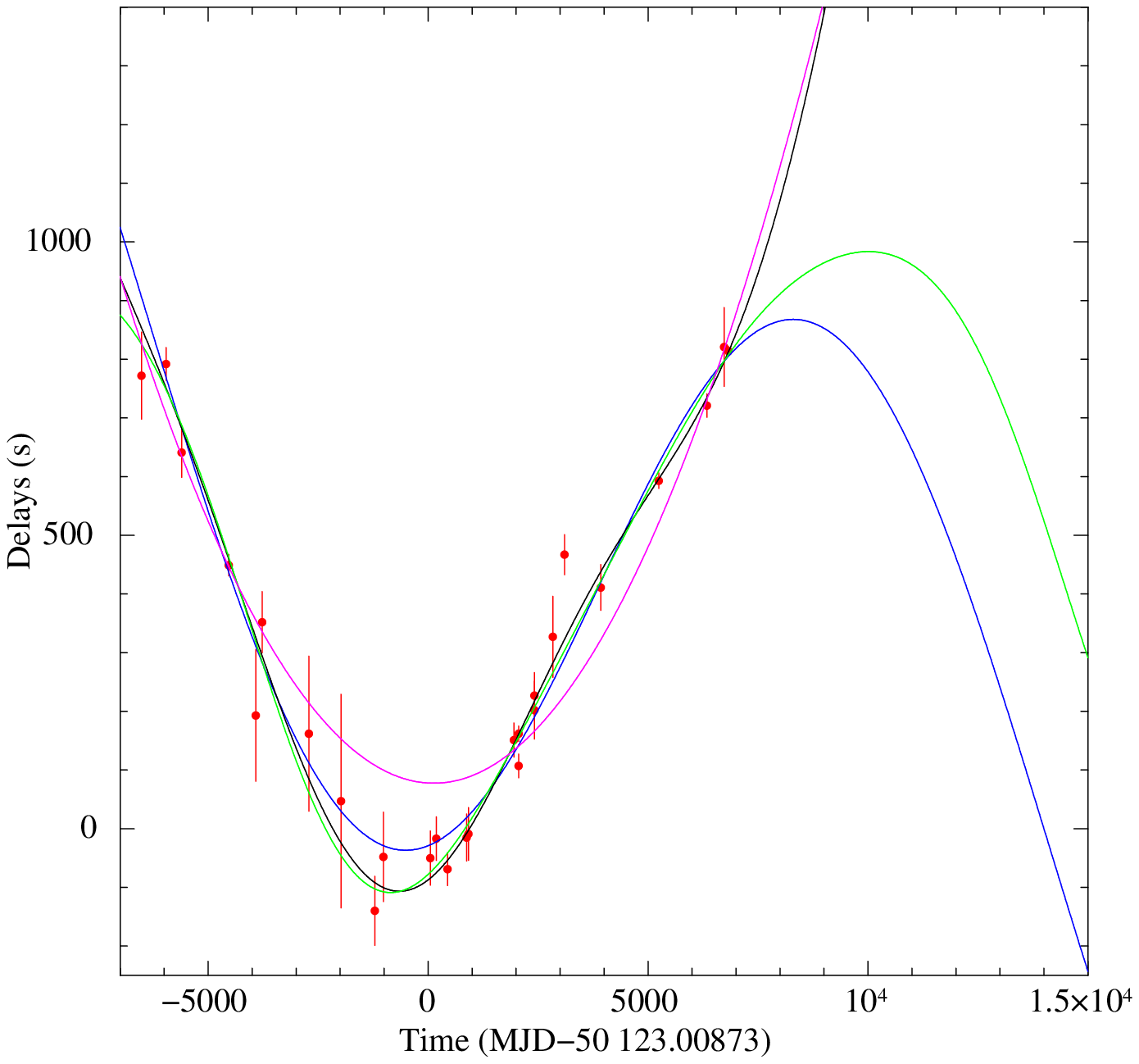}\hspace{0.3truecm}
\includegraphics[width=8.5cm]{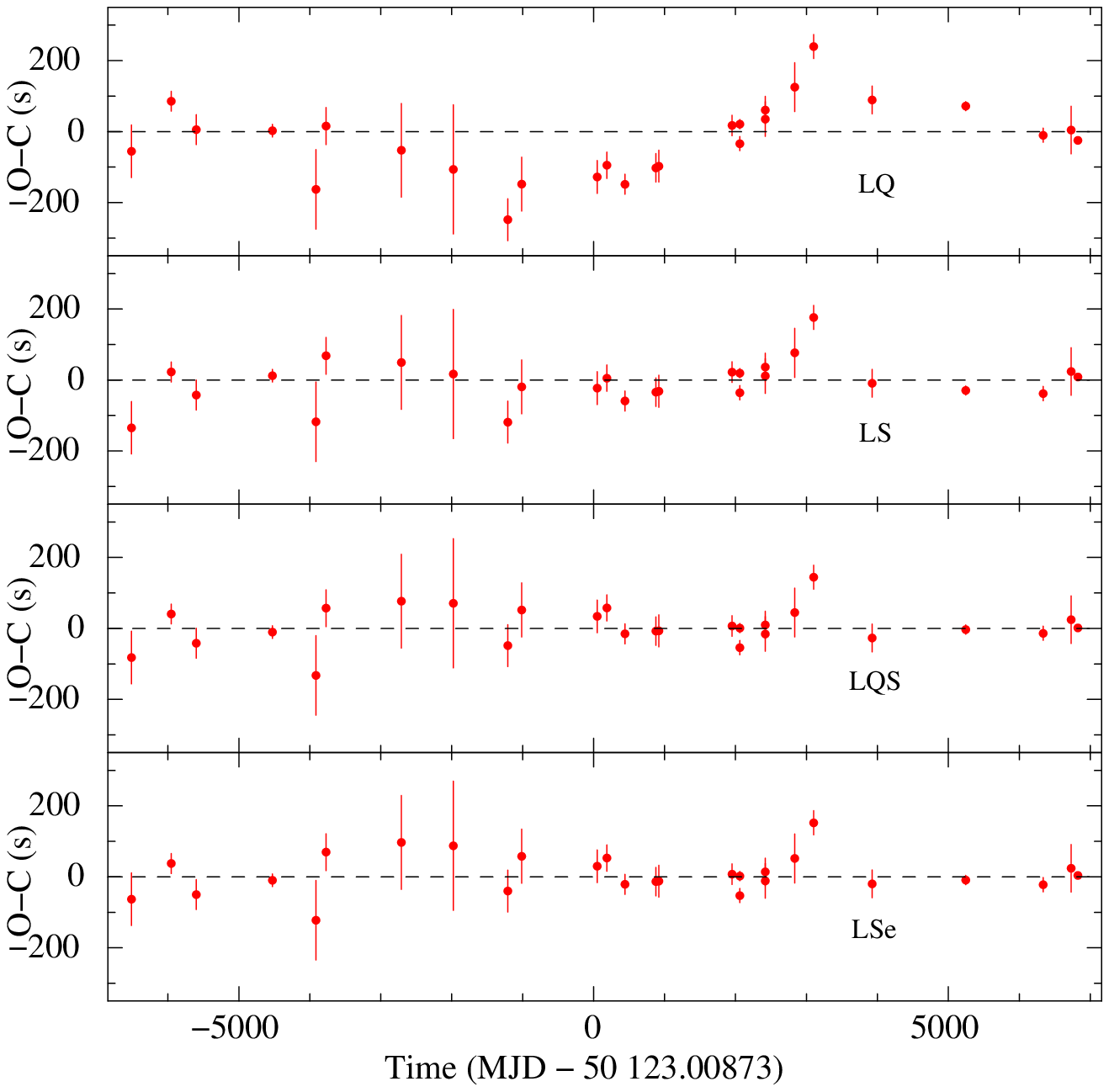}
\caption{Left panel: dips's arrival time delays versus time. The magenta,  blue, black , and  green
  curves are the best-fit curves obtained using the linear+quadratic
  (LQ), linear+sinusoidal (LS), linear+quadratic+sinusoidal (LQS),  and
  linear+sinusoidal function  taking into account a possible
  eccentricity (LSe), respectively. Right panel: observed minus calculated delays in units
of seconds.  The residuals, from the top to the bottom, correspond to
the LQC,  LS, LQS, and LSe function, respectively. }
 \label{delays}
\end{figure*}
The inferred delays, in units of seconds, of the dip arrival times
with respect to a constant 
orbital period are reported in Tab. \ref{Tab1}.
For each point we computed the corresponding cycle and the 
dip arrival time in days with respect to the adopted $T_0$. 
We show   the delays vs. time in Fig. \ref{delays} (left panel).

Initially we fitted the  delays with a quadratic function 
$$
y(t) = a+b t+ c t^2,
$$ 
where $t$ is the time in days with respect to $T_0$, $a=\Delta T_0$ is
the correction to $T_0$ in units of seconds, $b=\Delta P/P_0$ in units
of s d$^{-1}$  with $\Delta P$ the correction to the orbital period, and
finally, $c= 1/2 \;\dot{P}/P_0$ in units of s d$^{-2}$, with
$\dot{P}$, that is the orbital period derivative.  The quadratic form
does not fit the data, we obtained  $\chi^2({\rm {d.o.f.}})$ of
194.6(24).  Here, and in the following, we scaled the uncertainties in the
parameters  by a factor $\sqrt{\chi^2_{red}}$ to take
 a value of $\chi^2_{red}$ of the best-fit model larger than 1 into account.
The best-fit parameters are shown in the second column of  
Tab. \ref{fit_result}. The corresponding quadratic ephemeris
 (hereafter LQ ephemeris) is
\begin{equation}
\label{quadratic_eph}\begin{split}
T_{dip}(N) = {\rm MJD(TDB)}\; 50\,123.0096(3) + \frac{3\,000.65094(14)}{86\,400}  N +\\
+2.37(12) \times 10^{-13} N^2, 
\end{split}\end{equation}
where $N$ is the number of cycles, $50,\!123.0096(3)$ MJD is the new
Epoch of reference, the revised orbital period is
$P=3,\!000.65094(14)$ s, and the orbital period derivative obtained
from the quadratic term is $\dot{P} = 1.36(7) \times 10^{-11}$
s/s. The obtained quadratic ephemeris is compatible with that
reported by \cite{Hu_08}.   We show the best-fit curve in
  Fig. \ref{delays} (left panel) and the corresponding residuals in
  units of seconds in Fig. \ref{delays} (right panel, upper plot).
\begin{table*}
\setlength{\tabcolsep}{3pt} 
  \caption{Best-fit values of the parameters of the functions used to fit the 
delays.\label{fit_result}}
\begin{center}
\begin{tabular}{l c c c c c c c}          
\hline                                             
\hline  
Parameters & LQ  & LQC & LS & LQS & \multicolumn{3}{c}{LSe}\\  
\hline                                             
$a$   (s)      & $78 \pm 23$ &  $-2.7^{+2.1}_{-11.2}$    & $584\pm 157$ & $16 \pm 22$ &   $229 \pm 336$     &  $56 \pm 322$ &   $1.1 \pm 299.2$ \\
$b$   ($\times 10^{-3}$ s d$^{-1}$ )    & $-4 \pm 4$ & $37.1 \pm 0.4$ &  $-43 \pm 23$  & $-4 \pm 3$  &  $3 \pm 20$    &  $3 \pm 19$&  $5 \pm 22$ \\
$c$   ($\times 10^{-5}$ s d$^{-2}$)& $1.70 \pm 0.09$ &  $2.13 \pm
                                                       0.03$   &  -- &
                                                                       $1.79 \pm 0.09$ &-- & -- &--\\
$d$  ($\times 10^{-9}$ s d$^{-3}$) & -- & $-1.35 \pm 0.12$  & -- & --&-- & -- &--\\
$A$   (s)      &  -- &--  & $658 \pm 206$& $130 \pm 15$  &   $519 \pm 47$    &  $548 \pm 43$&   $577 \pm 43$ \\
$t_\phi$ (d)      &  --  & --  &  $3\,897 \pm 332$  &  $1\,356 \pm
                                                      203$ &
                                                             $-3\,723
                                                             \pm
                                                             1\,100$
               & $-3\,150 \pm 1\,116$&   $-2\,923 \pm 1\,034$ \\
$P_{mod}$  (d)  &  -- & --&  $20\,409 \pm 3\,381$&  $9\,302 \pm 752$ &
                                                                       17$\,$100
                                                                       (fixed)&
                                                                                18$\,$600 (fixed)& 20$\,$100 (fixed)\\
$\varpi$ (deg)   &  -- & --&--&--&  $195 \pm 26$  & $210 \pm 28$&
                                                                  $217
                                                                  \pm
                                                                  27$\\
$e$  &  -- & --&--&--&$0.26 \pm 0.20$      & $0.28 \pm 0.15$&  $0.32 \pm 0.13$  \\
$\chi^2$(d.o.f.) &194.6(24)& 92.4(23) &63.7(22)&39.4(21)&51.8(21)   &48.2(21)&45.8(21)\\

\hline                                             

\end{tabular}
\end{center} {\small \sc Note} \footnotesize--- The reported errors
are at 68\% confidence level.  The fit parameters of the delays
are obtained using LQ (column 2), LQC (column 3), LS (column 4),
LQS function (column 5), and LSe (columns 6, 7, and 8), respectively.
\\
\end{table*}
\begin{figure*}
\centering
\includegraphics[width=8.5cm]{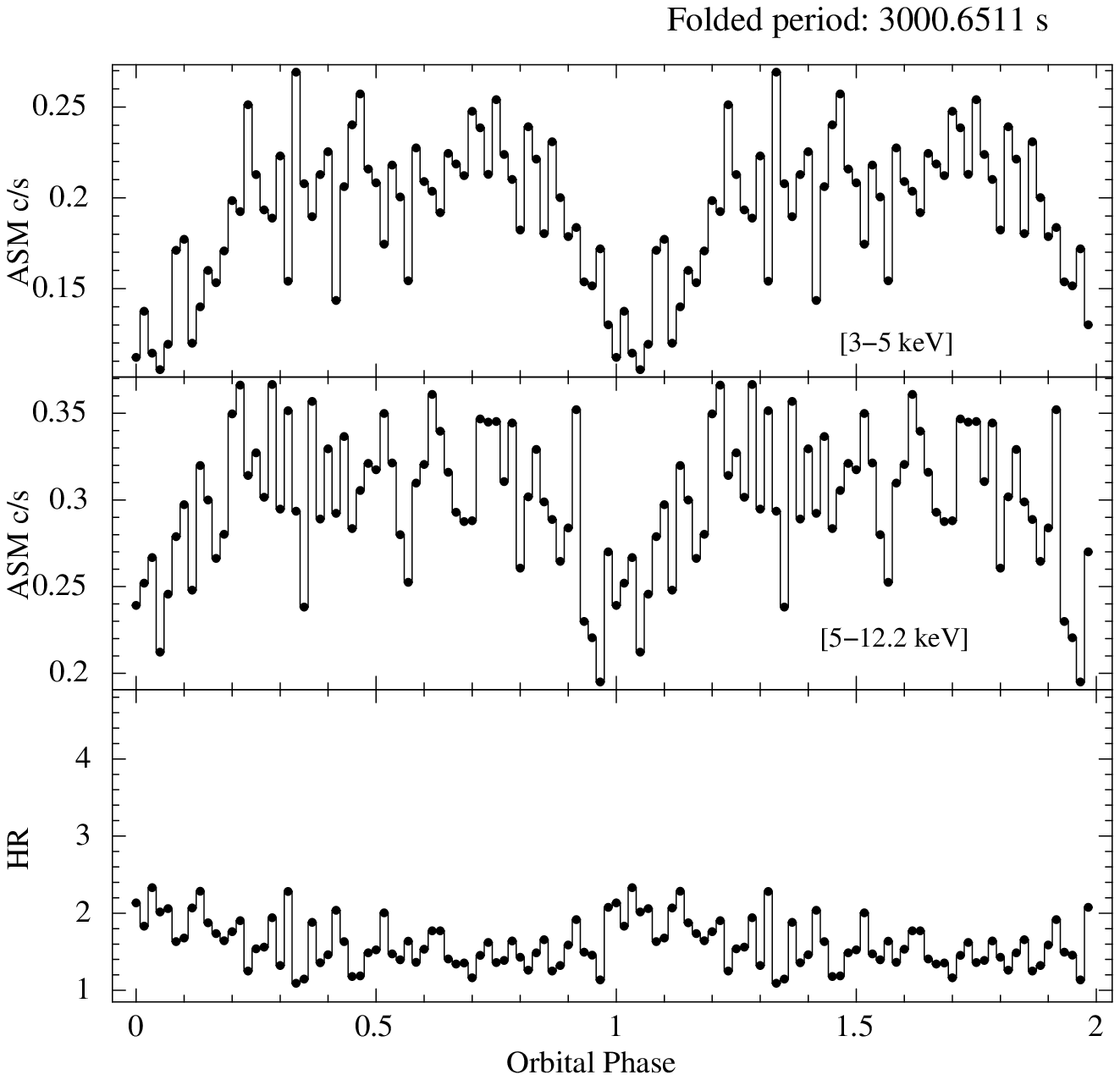}\hspace{0.3truecm}
\includegraphics[width=8.5cm]{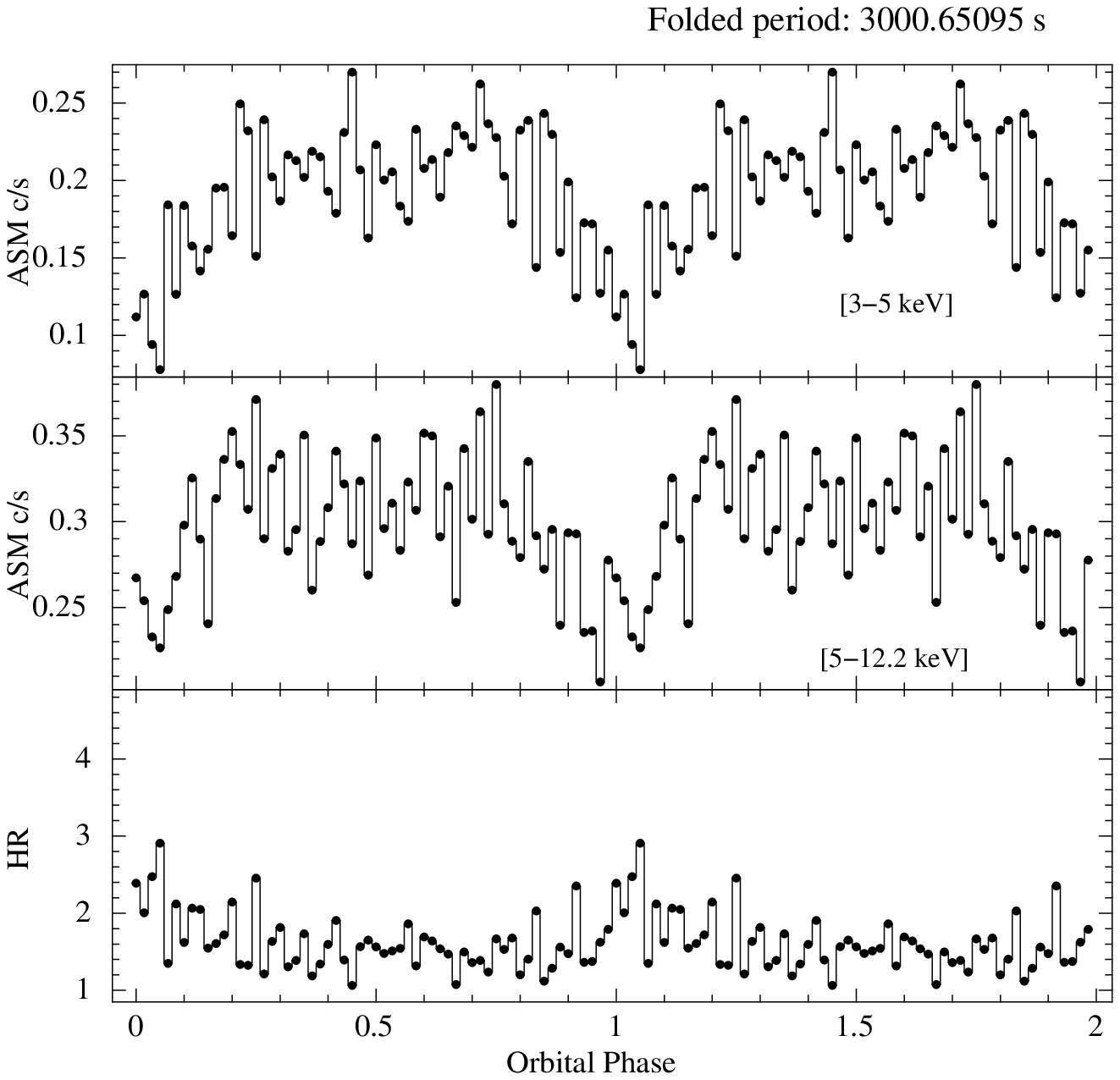}
\caption{Folded  RXTE/ASM light curve of XB 1916-053 in the 3-5 and
  5-12.2 keV energy range (top and middle panels). The corresponding
  hardness ratios (HRs) are plotted in the bottom panels. The left and
right plots show the folded  RXTE/ASM light curve using the ephemeris
discussed by  \cite{Hu_08} and  LQ ephemeris
(eq. \ref{quadratic_eph}) shown in the Sect. 3, respectively.  Each
phase-bin is about  50 s.}
 \label{asm_ratio_LQ}
\end{figure*}

As we obtained a large value of the  $\chi^2$,
 we fitted the delays vs. time adding a cubic
term to the previous parabolic function, i.e. 
$$
y(t) = a +bt+ct^2+dt^3,
$$
where $a$, $b$ and $c$ are above defined whilst the cubic term, $d$, is
defined as $\ddot{P}/(6 P_0),$ and  $\ddot{P}$ indicates the
temporal derivative of the orbital period derivative. Fitting with a
cubic function, we obtained a $\chi^2({\rm {d.o.f.}})$ of 92.4(23) with a
$\Delta \chi^2 $ of 101.2 and an F-test probability of chance
improvement of $4.2 \times 10^{-5}$ with respect to the quadratic
form.  The best-fit values are shown in the third column of Tab. \ref{fit_result}.
The corresponding  ephemeris  (hereafter LQC ephemeris) 
is
\begin{equation}
\label{cubic}\begin{split}
T_{dip}(N) = {\rm MJD(TDB)}\; 50\,123.00870^{+0.00005}_{-0.00026} + \\+\frac{3\,000.65239(3)}{86\,400}  N +
2.97(12) \times 10^{-13} N^2 -2.2(4) \times 10^{-22} N^3;
\end{split}\end{equation}
in this case we find an orbital period derivative of $1.71(7) \times
10^{-11}$ s/s and its derivative is $\ddot{P}=-3.8(0.7) \times 10^{-20}$
s/s$^2$. 

We also fitted the delays using a linear 
plus a sinusoidal function having the following terms
\begin{equation}
\label{delay_linear_sin_eph}\begin{split}
y(t) = a+b t+ A \sin\left[\frac{2 \pi}{P_{mod}}(t-t_\phi)\right],
\end{split}\end{equation}
where $a$ and $b$ are defined as above, $A$ is the amplitude of the
sinusoidal function in seconds, $P_{mod}$ is the period of the sine function in 
days, and, finally, $t_\phi$ is the time in days referred to $T_0$ at which 
the sinusoidal function is null.  We obtained a value of $\chi^2({\rm {d.o.f.}})$ of
63.7(22) with a $\Delta \chi^2 $ of  131 with respect to the quadratic form. 
The best-fit parameters are shown in the fourth column of  
Tab. \ref{fit_result}.  The best-fit  function is indicated 
with a blue curve in   Fig. \ref{delays} (left panel)
and the 
corresponding residuals are shown in Fig. \ref{delays} 
(right panel, the second plot from the top). The residuals are flatter than
those obtained in the previous case. Using the sinusoidal
function,  the dip time obtained from the OSO-8 observation is distant
 $\sim$200 s from the expected value.
The corresponding  ephemeris
 (hereafter LS ephemeris) is
\begin{equation}
\label{linear_sin_eph}\begin{split}
T_{dip}(N) = {\rm MJD(TDB)}\; 50\,123.01549(18) + \\+\frac{3\,000.6496(8)}{86\,400}  N +
A \sin\left[\frac{2 \pi}{N_{mod}} N-\phi\right], 
\end{split}\end{equation}
where $N_{mod}=P_{mod}/P_0 = 587\,659.53 \pm 97\,351.67$ 
 and $\phi = 2 \pi t_\phi/P_{mod}= $ with  $P_{mod} = 55.9 \pm 9.3$ yr.
\begin{figure*}
\centering
\includegraphics[width=8.5cm]{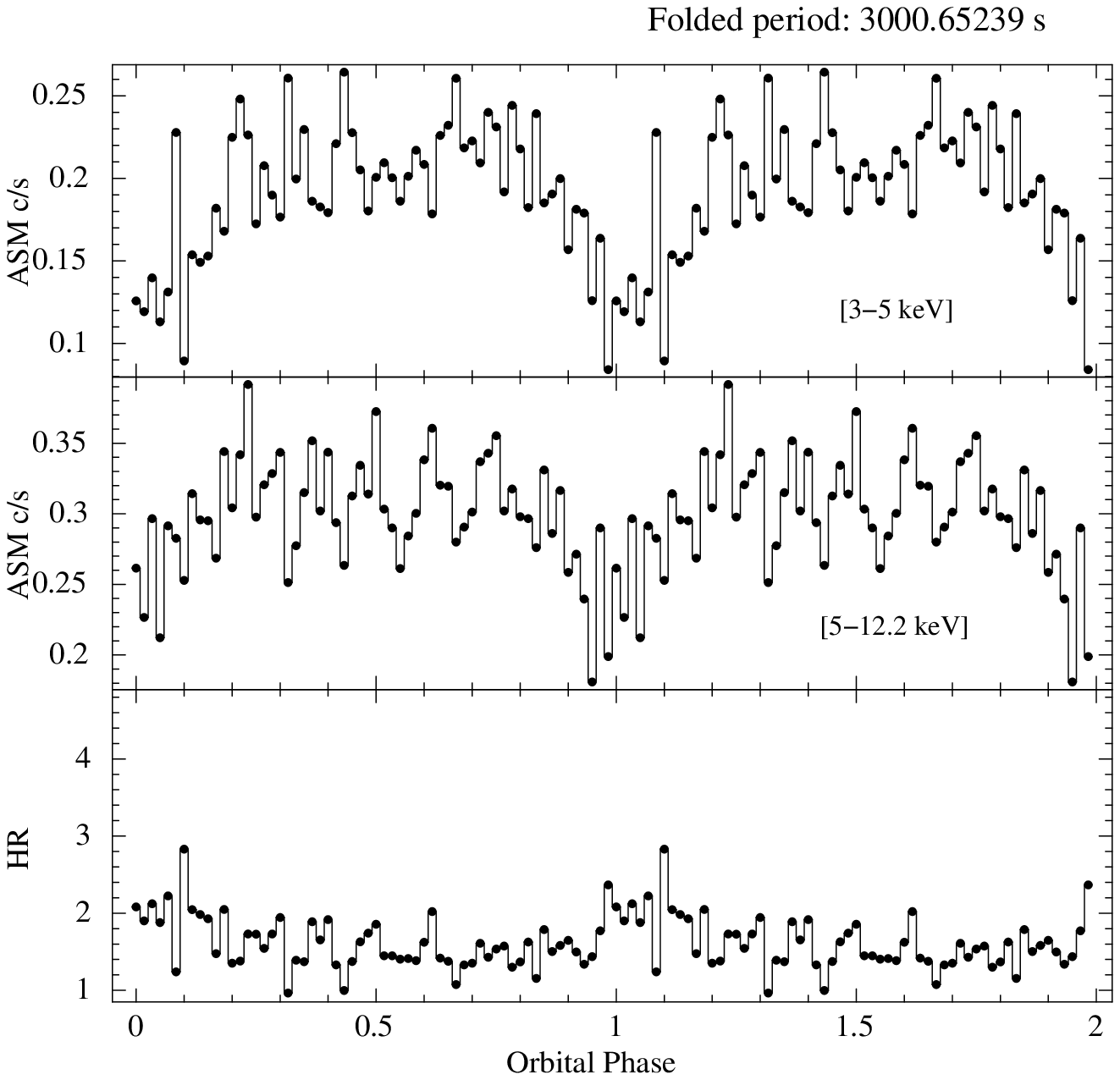}\hspace{0.3truecm}
\includegraphics[width=8.5cm]{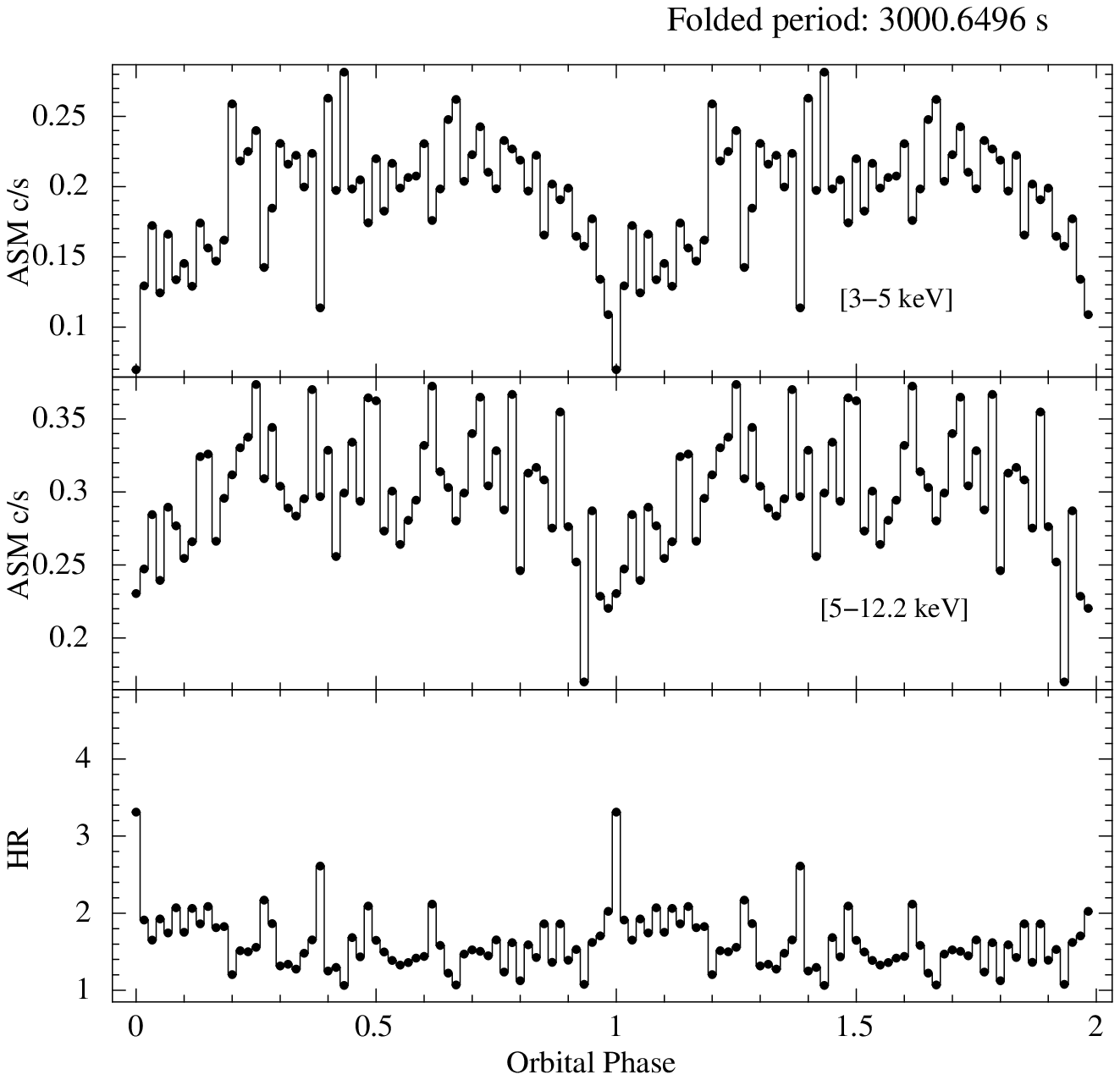}
\caption{Left and
right plots show the folded  RXTE/ASM light curve using  LQC ephemeris
(eq. \ref{cubic}) and  LS ephemeris
(eq. \ref{linear_sin_eph}), respectively.  Each phase-bin is about 50 s.}
 \label{asm_ratio_LQC}
\end{figure*}
\begin{figure*}
\centering
\includegraphics[width=8.5cm]{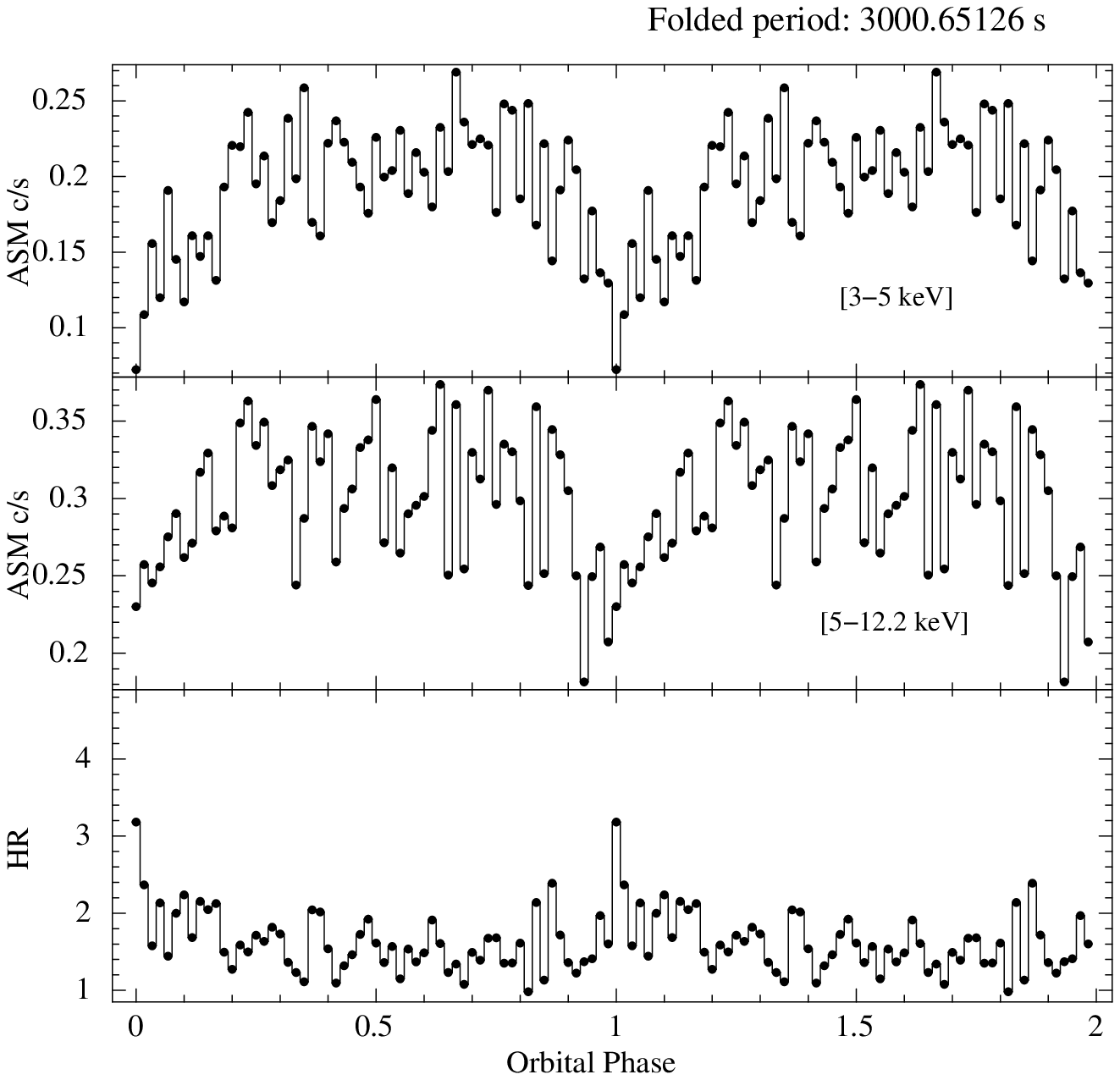}\hspace{0.3truecm}
\includegraphics[width=8.5cm]{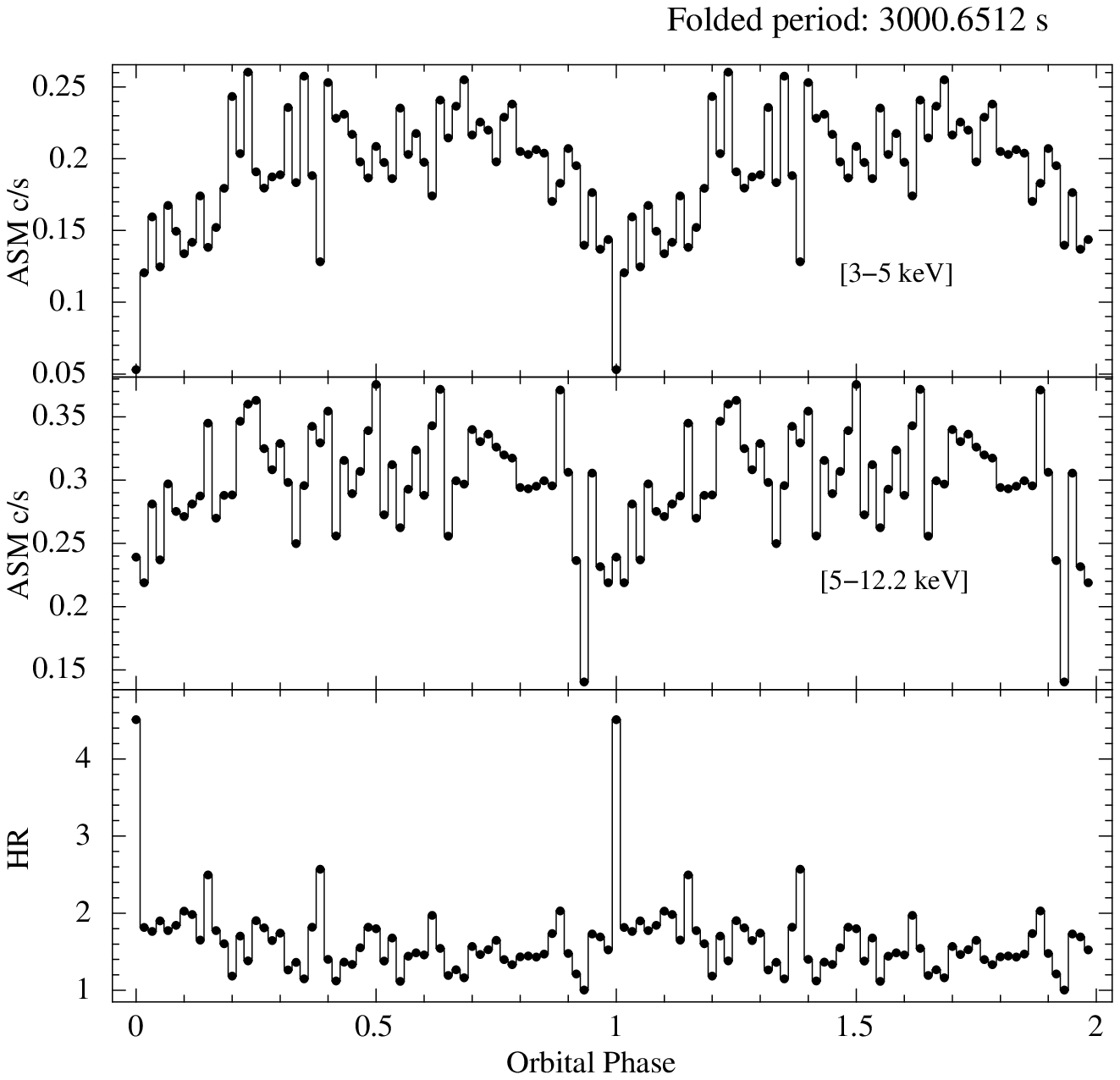}
\caption{Left and right plots show the folded RXTE/ASM light curve
  using LQS ephemeris (eq. \ref{linear_quad_sin_eph}) and LSe
  ephemeris (eq. \ref{linear_sin_e_eph}) with $P_{mod}=18\,600$ d,
  respectively.  Each phase-bin is about 50 s.}
 \label{asm_ratio_LQS}
\end{figure*}
This functional form significantly improves the fit, even though it does
not take the possible presence of an orbital period
derivative into account. 

We added a quadratic term to take  
the possible presence of an orbital period derivative and
 fitted the delays into account, using the function
\begin{equation}
\label{delay_linear_quad_sin_eph}\begin{split}
y(t) = a+b t+ c t^2 + A \sin\left[\frac{2 \pi}{P_{mod}}(t-t_\phi)\right].
\end{split}\end{equation}
We obtained a value of $\chi^2({\rm {d.o.f.}})$ of
39.4(21) and a F-test probability of chance improvement 
with respect to the LS ephemeris of $1.7 \times 10^{-3}$. 
The best-fit parameters are shown in the fifth column of  
Tab. \ref{fit_result}.  The best-fit  function is indicated 
with a black curve in Fig. \ref{delays} (left panel) and the 
corresponding residuals are shown in Fig. \ref{delays} (right panel, the third plot from the top).
The corresponding  linear+quadratic+sinusoidal ephemeris 
(hereafter LQS ephemeris) is
\begin{equation}
\label{linear_quad_sin_eph}\begin{split}
T_{dip}(N) = {\rm MJD(TDB)}\; 50\,123.0089(3) + \frac{3\,000.65126(10)}{86\,400}  N +\\
+2.50(12) \times 10^{-13} N^2 + A \sin\left[\frac{2 \pi}{N_{mod}} N-\phi\right], 
\end{split}\end{equation}
with $N_{mod} = 267\,837.87  \pm 21\,652.90  $ and $\phi = 0.92 \pm 0.16$. 
The corresponding orbital period derivative is $\dot{P} = 1.44(7) \times 
10^{-11}$ s/s and the period of the  modulation is $P_{mod} = 25.5 \pm 2.1$
yr.  

Our analysis of the delays suggests  that a quadratic or a quadratic plus a cubic
term do not fit the delays. A better fit is obtained using a
sinusoidal function with a period close to 20$\,$000 d and, finally,
adopting a sinusoidal plus a quadratic term, we obtain the best fit of
the delays. In this latter case, the sinusoidal function has a period
of  9$\,$300 d,  about half of that obtained using only the
sinusoidal function. Moreover, the orbital period derivative  $\dot{P} = 1.44(7) \times 
10^{-11}$ s/s  \citep[compatible with $\dot{P} = 1.5(3) \times
10^{-11}$ s/s obtained by][]{Hu_08}
is extremely high to be explained by  a conservative mass transfer and   
loss of angular momentum from the binary system  for gravitational
radiation (see next section). This awkward result can be bypassed if
the quadratic term is merely an approximation of a further sinusoidal
function with a larger orbital period with respect to   9$\,$300 d.

 Under this assumption, the best fit obtained using the LQS ephemeris
could be explained using a different scenario, where the quadratic
term mimics the fundamental harmonic of a series expansion whilst
the sinusoidal term is the first harmonic. This seems also suggested
by the best fit obtained using  the LS function
(eq. \ref{delay_linear_sin_eph}), since  we obtain 
a modulation period, which is almost twice that obtained using the LQS function
(eq. \ref{delay_linear_quad_sin_eph}). 

If  we assume that XB 1916-053 is part of a hierarchical triple system
then the measured delays are also affected by the influence of a third
body. If the orbits  of the third
body and of the X-ray binary system around the common centre of mass are 
slightly elliptical then the delay $\Delta_{DS}(t)$ associated with the Doppler
shift can be expressed as
\begin{equation}
\begin{split}
\label{delta}
\Delta_{DS}(t) & =A \biggl\{\sin(m_t+\varpi)+   \frac{e}{2}
\left[\sin(2m_t+\varpi) -3\sin(\varpi) \right] +\\ &
+\frac{e^2}{4}[     2  \sin(3m_t+\varpi)
  -\sin(m_t+\varpi)\cos(2m_t+1) +\\ & -2\sin(m_t)\cos(\varpi)]  \biggr\},
\end{split}
\end{equation}
where 
$$
m_t=\frac{2\pi}{P_{mod}}(t-t_{\phi})
$$
is the mean anomaly;
$e$ is the eccentricity of the orbit; $P_{mod}$ is the orbital
period of both the X-ray binary system and the third body around the
common centre of mass; $\varpi$ denotes
the periastron angle; $t_{\phi}$ is
the passage time at the periastron;\ and $A=a \sin i/c$ is the
projected semi-major axis of the orbit, described by the centre of mass
of the X-ray binary system around the centre of mass of the triple
system. We neglect third and higher order terms in Eq. \ref{delta}.
Limiting Eq. \ref{delta} to the first-order terms, it becomes the
expression shown by \cite{vdk_84}.
\begin{figure*}
\centering
\includegraphics[width=8.5cm]{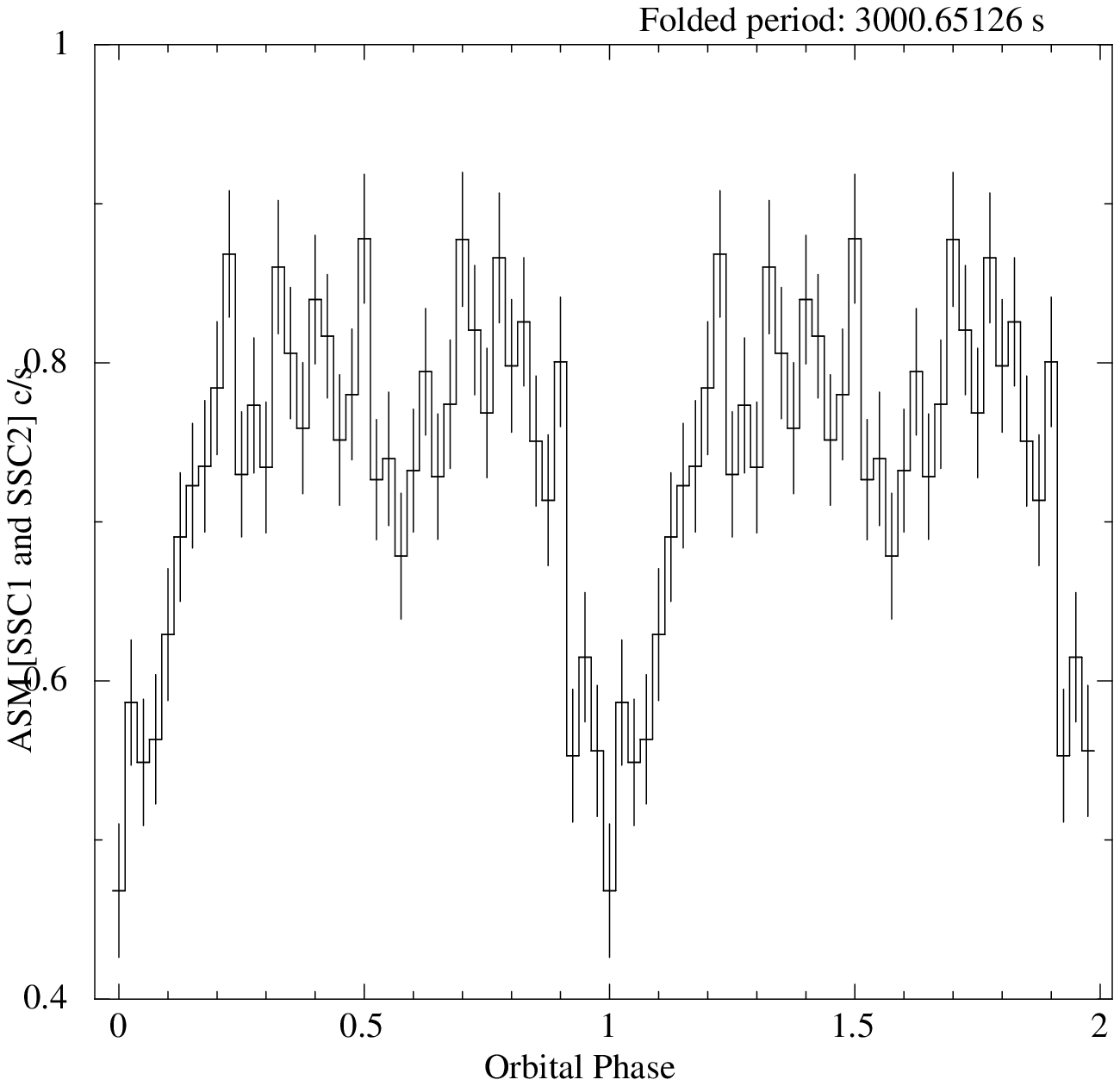}\hspace{0.3truecm}
\includegraphics[width=8.5cm]{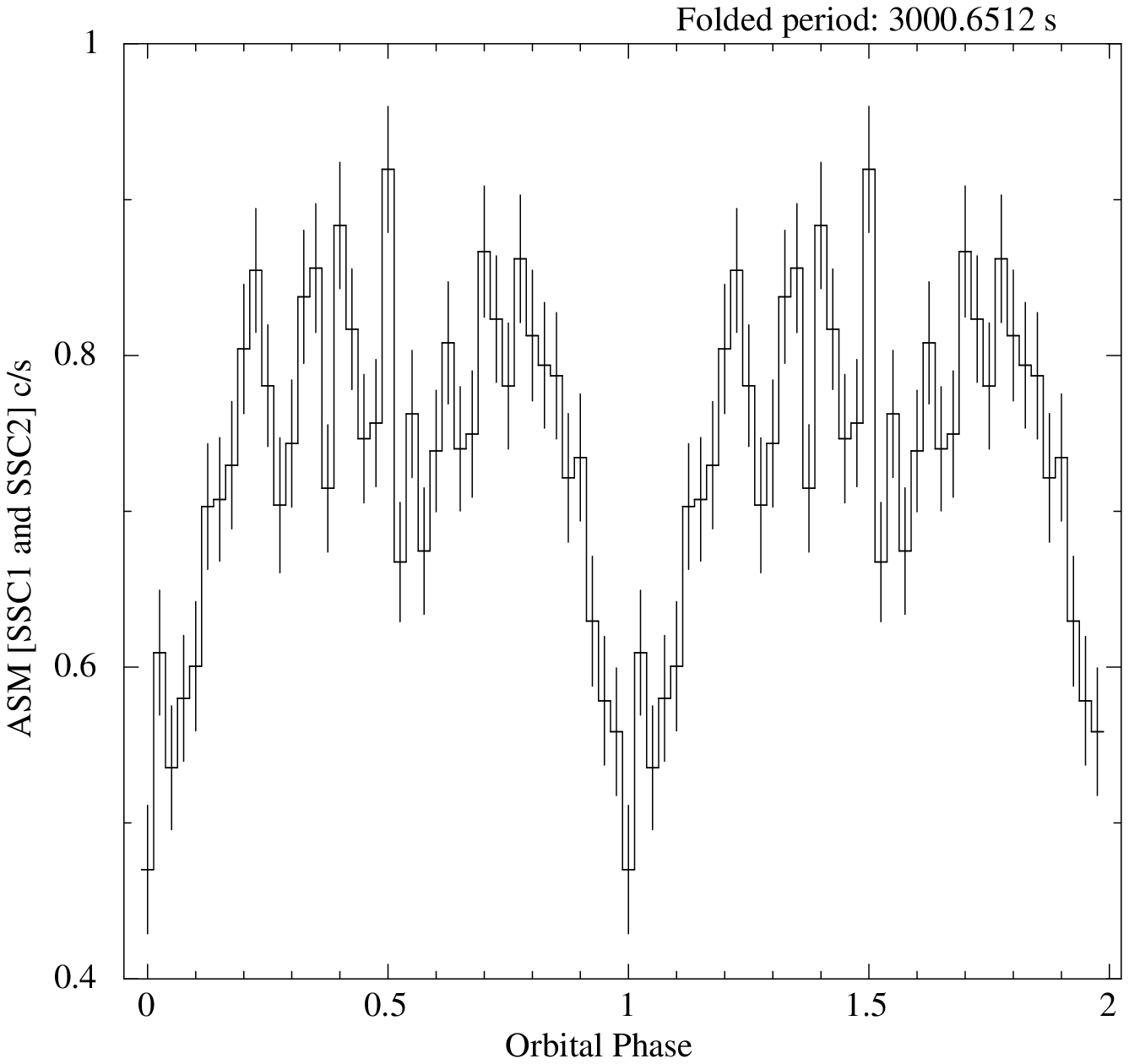}
\caption{Folded RXTE/ASM light curve of XB 1916-053 selecting the
    events from SSC1 and SSC2. No energy filter is applied. Each
    phase-bin correspond to 75 s.  Left panel:  folded light curve using the
  LQS ephemeris (eq. \ref{linear_quad_sin_eph}).  Right panel: folded light curve  using the
  LSe ephemeris (eq. \ref{linear_sin_e_eph}). }
 \label{ASM_LSe}
\end{figure*}
Then, we fitted the delays using
$$
y(t) = a +bt+\Delta_{DS}(t).
$$
Because the 27 available points do not cover a whole period, we
arbitrarily fixed
the value of $P_{mod}$ at 18$\,$600, 17$\,$100, and 20$\,$100 d, which  are
the best, lower, and upper  value of the period obtained  from the LQS
ephemeris multiplied by  a factor of two. The best-fit parameters are
shown in Tab. \ref{fit_result} (columns 6, 7, and 8). The $\chi^2({\rm {d.o.f.}})$ are
similar for 
the three adopted periods and the F-test probability of chance
improvement with respect to LS function is  $4.1 \times 10^{-2}$,  $1.7 \times 10^{-2}$,  and
$0.9 \times 10^{-2}$  for a  $P_{mod}$ value of  17$\,$100, 18$\,$600, and 20$\,$100
d, respectively. In the following, we discuss
the case of $P_{mod}=18\,600$ d.  The best-fit  function is indicated 
with a green curve in  Fig. \ref{delays} (left panel). The 
corresponding residuals are shown in Fig. \ref{delays} 
(right panel, lower plot). The corresponding ephemeris (hereafter LSe
ephemeris) is
\begin{equation}
\label{linear_sin_e_eph}\begin{split}
T_{dip}(N) =  {\rm MJD(TDB)}\; 50\,123.010(3)  +
\frac{3\,000.6512(6)}{86\,400}  N +\\+\Delta_{DS}(N).
\end{split}\end{equation}

To verify the robustness of our results, we 
produced the folded light curves in the 3-5 and 5-12.2 keV
energy bands of XB 1916-053 obtained from the All Sky Monitor (ASM) on
board RXTE  using the ephemerides shown above. We inferred those ephemerides 
 using only pointing observations from which we obtained 27 points spanning from 1978 to 
2014, whilst  the RXTE/ASM light curves
cover  from  1996 Sep 01 to 2011 Oct 31. 
 We applied the barycentre corrections to the RXTE/ASM events. As a
first step, we folded the RXTE/ASM light curves of XB 1916-053 using
the LQ ephemeris reported by \cite{Hu_08} and by us
(Eq. \ref{quadratic_eph}), adopting 60 phase-bins per period corresponding to $\sim$50 s per bin. The folded light curves and the
corresponding hardness ratios (HRs) are shown in
Fig. \ref{asm_ratio_LQ}.  None of the  HR  show an evident increase
at phase zero as we would expect if the ephemerides well define the dip
arrival times. This implies that those ephemerides do not correctly
predict the dip arrival times contained in the RXTE/ASM light curve.
 Adopting the LQC ephemeris (eq.  \ref{cubic}), the
maximum value of HR (that is 2.8) is reached at phase 0.1 (see
Fig. \ref{asm_ratio_LQC}, left panels). Also in this case, the LQC
ephemeris does not predict the dip arrival times in the ASM
light curves of XB 1916-053.  Using the LS ephemeris
(Eq. \ref{linear_sin_eph}) to fold the light curves, we obtained that
the maximum value of HR is reached at phase zero and is close to
3.4 (see Fig. \ref{asm_ratio_LQC}, right panels). In contrast, with the LQS
ephemeris (Eq. \ref{linear_quad_sin_eph}) the maximum value of the HR
falls in two phase-bins close to phase zero (see
Fig. \ref{asm_ratio_LQS}, left panels) and the maximum value of HR is
3.2, which  is smaller than the value obtained with the LS ephemeris.
Finally, we folded the RXTE/ASM light curves using the LSe ephemeris
(eq.  \ref{linear_sin_e_eph}).  We show the folded light curves and the
corresponding HR in Fig. \ref{asm_ratio_LQS} (right panel). In this
last case the maximum value of the HR falls in only one phase bin at
phase zero and the maximum value of the HR is about 4.5.

We also folded the RXTE/ASM light curve  (not
filtered in energy) using 
the LQS and LSe ephemerides once we selected 
the  events from the 
Scanning Shadow Cameras (SSCs) 1 and 2. Adopting 40 phase-bins per
period (that is each bin is 75 s), the folded light curves are very
similar (see Fig. \ref{ASM_LSe}), indicating that the two ephemerides are
statistically equivalent.  The dip is clearly observed at phase zero, the 
ASM count rate is reduced during the dip of 60\% with respect  to
the
persistent
emission. Finally, the goodness of the two ephemerides allows us to observe
the presence of a secondary dip at phase 0.55, which is typically observed in
several dipping sources \citep[see][for XB 1916-053]{Grindlay_89}.

\section{Discussion}

From the study of the 27 dip arrival times obtained from the pointed
observations of XB 1916-053 and of the RXTE/ASM light curves, we find
that the quadratic and cubic ephemerides do not correctly predict  the dip arrival
times on a long time span; whilst to well fit the delays, we need to
use  a function that contains at least  linear and  sinusoidal terms
 (LS ephemeris, see
Eq. \ref{linear_sin_eph}).  The addition of a quadratic term to the LS
ephemeris (Eq. \ref{linear_quad_sin_eph}) gives a probability of
chance  improvement obtained with a F-test of $1.7 \times
10^{-3}$   with respect to the LS ephemeris. Finally, using the ephemeris shown in
Eq. \ref{linear_sin_e_eph}, the probability of chance improvement,  also with respect to
  the LS ephemeris, is  $1.7 \times 10^{-2}$.
The LQS and LSe ephemerides paint  two different physical scenarios for XB 1916-053.
In the first case the orbital period derivative of the X-ray binary
system is $\dot{P} = 1.44(7) \times 10^{-11}$ s/s and the observed
delays associated with the dip arrival times are affected by a
  relatively low-amplitude ($\sim 130$ s) sinusoidal modulation 
with a period close to 26 yr. 
In the second case the
orbital period derivative is fixed to zero and the 
modulation of the delays is solely sinusoidal with an amplitude of
$\sim 550 $ s and an orbital period close to 51 yr. 
We  explain in the following the sinusoidal modulation for both the scenarios, assuming
the presence of a third body forming a hierarchical triple system with
XB 1916-053, which alters the observed dip arrival times. 
 
We start by discussing the plausible values of the companion star mass $M_2$.
We know that the companion star is a
degenerate star and its radius $R_2$ has to be equal to its Roche lobe
radius $R_{L2}$ since the binary system is in the Roche lobe overflow
(RLOF) regime.  Rearranging the Eq. 3.3.15 in \cite{Shapiro_83}, the
mass-radius relation for a degenerate star can be written as
$$
\frac{R_2}{R_{\odot}}=0.04\left(\frac{Z}{A}\right)^{5/3}\left(\frac{M_2}{M_{\odot}}\right)^{-1/3}
= 0.0126\; (1+X)^{5/3}m_2^{-1/3},
$$ 
where $Z$ and $A$ are the  atomic number and the  
atomic weight of
the matter composing the star, and where we assumed that the matter is
only composed of hydrogen and helium.   The factor $Z/A$  
is the average of Z/A for matter composed of hydrogen and  helium, 
$X$ is the fraction of hydrogen in the star and, finally,
$m_2$ is the companion star mass in units of solar mass.
This equation has to be corrected for  the thermal bloating factor  $f,$ which is the ratio
of the companion star radius to the radius of a star with the same mass
and composition, that is completely degenerate and supported only by 
the Fermi pressure of the electrons; then  the factor $f$ is $>1$.
The Roche lobe radius of the companion star can be written as
$$
R_{L2}=0.46224\;a\;\left(\frac{m_2}{m_1+m_2}\right)^{1/3},
$$
where  $a$ is the orbital separation of the binary system and $m_1$ is 
the neutron star (NS) mass in unit of solar mass. We can
write  $a$  in terms of the orbital period $P$, $m_1$, and $m_2$, using  Kepler's third law.
Combining the last two equations and Kepler's third law,
we obtain 
\begin{equation}
\label{mass}
m_2=0.0151 \; (1+X)^{5/2} f^{3/2}.
\end{equation}
\cite{nelemans}, analysing the optical
spectrum with the European Southern Observatory Very Large Telescope,
detected  a He-dominated accretion disc spectrum and suggested  
direct evidence for a helium donor. The authors  found a good match with an
LTE model consisting of pure helium plus overabundant
nitrogen. 
For this reason,  we assume $X=0$ in the rest
of the discussion.

The bolometric X-ray flux of XB 1916-053 was estimated by several
authors.  \cite{Galloway_08}, analysing a RXTE/PCA observation
of XB 1916-053, determined a persistent flux in the 2.5-25 keV of
$(3.82 \pm 0.04) \times 10^{-10}$ erg s$^{-1}$ cm$^{-2}$. The authors
corrected the flux for a bolometric factor $c_{bol} = 1.37 \pm 0.09$
to estimate the bolometric flux in the 0.1-200 keV energy range,
obtaining a bolometric flux of $(5.2 \pm 0.3) \times 10^{-10}$ erg
s$^{-1}$ cm$^{-2}$.  Recently, \cite{zhang_2014}, analysing a Suzaku
observation of XB 1916-053,  found a value of $F_{bol}$ in the
0.1-200 keV energy range between $5.5 \times 10^{-10}$ and
$6.1 \times 10^{-10}$ erg s$^{-1}$ cm$^{-2}$.  Finally, analysing the
persistent emission of the source during a BeppoSAX observation,
\cite{church_98} estimated a value of $F_{bol}$ in the 0.5-200 keV
energy range of $6.2 \times 10^{-10}$ erg s$^{-1}$ cm$^{-2}$.  Since
the RXTE/ASM light curve of XB 1916-053 shows that the count rate of
the source is almost constant over more than ten years, we adopt a conservative value for the
bolometric flux of $(5.5 \pm 0.5) \times 10^{-10}$ erg s$^{-1}$
cm$^{-2}$.

The distance $d$ to the source was estimated by   \cite{Galloway_08}
measuring the peak flux during the photospheric
radius expansion (PRE) in type-I X-ray bursts. Equation  8
in \cite{Galloway_08} can be rewritten
\begin{equation}
\label{eq_d}
\begin{split}
d= 8.32 \left(\frac{F_{pk,PRE}}{3 \times 10^{-8}{\rm erg\; s^{-1}\;cm^{-2}}}\right)^{-1/2}
m_1^{1/2} \left(1-0.296\;\frac{m_1}{r_{PRE}}\right)^{1/4}\\(1+X)^{-1/2} {\rm kpc},
\end{split}
\end{equation}
where $r_{PRE}$ is the
photospheric radius of the neutron star in units of 10 km and
$F_{pk,PRE}$ is the flux at the peak of the type-I X-ray burst during
the PRE. The authors measured
$F_{pk,PRE} =(2.9 \pm 0.4)\times 10^{-8} \; {\rm erg\;
  s^{-1}\;cm^{-2}}$
and $r_{PRE}\simeq 1.1$ for XB 1916-053 and concluded that the distance
to the source is $d=8.9 \pm 1.3$ kpc (adopting $X=0$) for a NS mass of
1.4 M$_{\odot}$.  The X-ray luminosity can be expressed as
$L_x=4 \pi d^2 F_{bol}$, where we roughly assume that the emitted flux
is isotropic. We obtain $L_x \simeq 5.2 \times 10^{36}$ erg s$^{-1}$
for a NS mass of 1.4 M$_{\odot}$, whilst we find
$L_x \simeq 6.6 \times 10^{36}$ erg s$^{-1}$ for a massive NS of 2.2
M$_{\odot}$.

\cite{rappa_87}  predicted the X-ray luminosity for highly compact binary systems
under the reasonable hypothesis that the main mechanism to lose angular momentum
is  gravitational radiation. Combining the Eqs. 8 and 13 in their work, we
obtain
\begin{equation}
\label{lx_rappa}
L_x \simeq \frac{5.2 \times 10^{42}}{1-1.5 \alpha (1-\beta)} m_1^{5/3}
P_m^{-14/3}(1+X)^5 \beta \eta f^3 \;{\rm erg \; s}^{-1},
\end{equation}
where $P_m$ is the orbital period in units of minutes, $\beta$ is the
fraction of matter yielded by the companion star and accreted onto the
NS, $\eta$ is the efficiency for converting gravitational potential
energy into X-ray emission, and $\alpha$ is the specific angular
momentum carried away by the mass lost from the system, in units of
$2 \pi a^2/P_{orb}$, where $a$ is the orbital separation
\citep[see][]{rappa_82}.  In
Eq. \ref{lx_rappa} we assume that the NS radius is 10 km.  Using the orbital period value of 3$\;$000.65 s,
assuming $\eta=1$ and a conservative mass transfer scenario
($\beta=1$), we find that $L_X \simeq 1.1 \times 10^{35} f^3$ erg s$^{-1}$
and $L_X \simeq 2.3 \times 10^{35} f^3$ erg s$^{-1}$ for a NS mass of 1.4
M$_{\odot}$ and 2.2 M$_{\odot}$, respectively.  Comparing the observed
luminosity and the predicted luminosity, we estimate that $f=3.6 \pm 0.4$ and
$f=3.0 \pm 0.3$ for a NS mass of 1.4 M$_{\odot}$ and 2.2 M$_{\odot}$,
respectively.  Substituting the obtained values of $f$ in
eq. \ref{mass}, we obtain that the companion star mass is
$M_2=0.10 \pm 0.02 $ M$_{\odot}$ and $M_2=0.078 \pm 0.012 $
M$_{\odot}$ for a NS mass of 1.4 M$_{\odot}$ and 2.2 M$_{\odot}$,
respectively.  The mass ratio $q=M_2/M_1$ of XB 1916-053 is between
$0.036 \pm 0.009$ and $0.071 \pm 0.009$.

\cite{Hu_08} inferred  the mass ratio of XB 1916-053 from the negative super-hump
period and found $q \simeq 0.045,$ which is compatible with
our estimated range of  values of $q$.  \cite{Chou_01} estimated a
value of $q \simeq 0.022$ using the period of the apsidal precession of the
accretion  disc of $P_{prec}=3.9087(8)$ d.  The value of $q$ obtained by
\cite{Chou_01} is outside the range that we find.

To estimate the orbital period derivative we use the eq. 11 shown in
\cite{rappa_87} that we rewrite as
\begin{equation}
\label{pdot_rappa}
\dot{P} \simeq \frac{1.54 \times 10^{-9}}{1-1.5 \alpha (1-\beta)} m_1^{2/3}
P_m^{-8/3}(1+X)^{5/2}  f^{3/2} \;{\rm s \; s}^{-1}.
\end{equation}
Using the value of $\dot{P} \sim 1.44 \times 10^{-11}$ s s$^{-1}$ (LQS
ephemeris) and
the orbital period value of 3$\;$000.65 s, we find that the thermal
bloating factor $f$ is 40 and 32 for a NS mass of 1.4 and 2.2 
M$_{\odot}$.  These values of $f$ are not physically plausible and
suggest that, in a conservative mass transfer scenario, 
 the value of the orbital period derivative cannot be that obtained
from the LQS ephemeris.

On the other hand, adopting an orbital period  of 3$\;$000.65 s 
and a factor $f$ of 3.6 and 3.0 for a NS mass of 1.4 and 2.2 M$_{\odot}$
we find  
$\dot{P} = (3.9 \pm 0.2) \times 10^{-13} $ s s$^{-1}$ and 
$\dot{P} = (3.98 \pm 0.15) \times 10^{-13} $ s s$^{-1}$ 
for a NS mass of 1.4  M$_{\odot}$  and 2.2  M$_{\odot}$,
respectively.  The orbital period derivative normalised to the orbital
period is $\dot{P}/P \simeq 4.2 \times 10^{-9}$ yr$^{-1}$ and weakly
depends on the NS mass.
We conclude that the conservative mass transfer scenario  with a thermal
bloating factor of the companion star  between three and four allows
us to explain the discrepancy between the  predicted and observed
X-ray luminosity, but  it does not solve the discrepancy between the
predicted and measured orbital period derivative obtained from the
LQS ephemeris. For this reason, we investigate the non-conservative
mass transfer scenario. 

Combining the eqs. \ref{lx_rappa} and \ref{pdot_rappa}, we obtain
 \begin{equation}
\label{ratio_rappa}
\frac{L_x}{\dot{P}} \simeq  3.38 \times 10^{51} m_1
P_m^{-2} \; \beta  f^{3/2} \eta   \;{\rm erg \; s}^{-1}.
\end{equation}
Adopting $L_x \simeq 5.2 \times 10^{36}$ erg s$^{-1}$, $\dot{P} = 1.44
\times 10^{-11}$ s s$^{-1}$, 
$P=3\;000.65$ s and fixing $\eta=1,$ we find that
$\beta f^{3/2} = 0.191$ for a NS mass of 1.4 M$_{\odot}$. Since $f >1,$
we expect that more than  81\% of the mass yielded by the companion
star leaves the system. Furthermore, since the measured values of
$L_x$ and $\dot{P}$ are positive, the term $1-1.5\alpha(1-\beta)$
in eqs.  \ref{lx_rappa} and \ref{pdot_rappa} should be positive. Solving for
 $\alpha$ while taking $\beta<0.191,$ we obtain that $\alpha < 0.823$. Because $\alpha$ is in
unit of $2 \pi a^2/P_{orb}$, we find that the matter should leave the
binary system from a distance $\bar{d}$ from the neutron star of
$\bar{d}<\alpha^{1/2} a$; the point of ejection in unit of orbital
separation is $\bar{x}=\bar{d}/a<\alpha^{1/2}$.  In the rest of the
discussion, we assume that the matter is ejected at the inner
Lagrangian point $x_{L1}$ of the binary system.  We rewrite the
eq. \ref{lx_rappa} as function of $f$ using the condition 
$\beta f^{3/2} = 0.191$. We find
 \begin{equation}
\label{lx_rappa_f}
L_x \simeq \frac{5.2 \times 10^{42}}{1-1.5 \;\bar{x_{L1}}^2 \;(1-0.191\;f^{-3/2})} m_1^{5/3}
P_m^{-14/3} 0.191 f^{3/2} \;{\rm erg \; s}^{-1},
\end{equation}
where $\bar{x_{L1}}$ is the position of the inner Lagrangian point in
units of orbital separation.  Using  eq. \ref {mass} and a NS
 mass of 1.4 M$_{\odot}$,  $\bar{x_{L1}}$ can be written as a 
cubic function of $f$ for values of the thermal bloating factor  between 1 and 10. We find
$$
\bar{x_{L1}}=0.915 -6.87 \times 10^{-2} f + 6.61 \times 10^{-3} f^2 -
2.88   \times 10^{-4} f^3,
$$
with an accuracy of $2 \times 10^{-3}$.
Combining the last equation and eq. \ref{lx_rappa_f}, we infer the
luminosity as function of $f$.  
\begin{figure}
\centering
\includegraphics[width=6.cm,angle=-90]{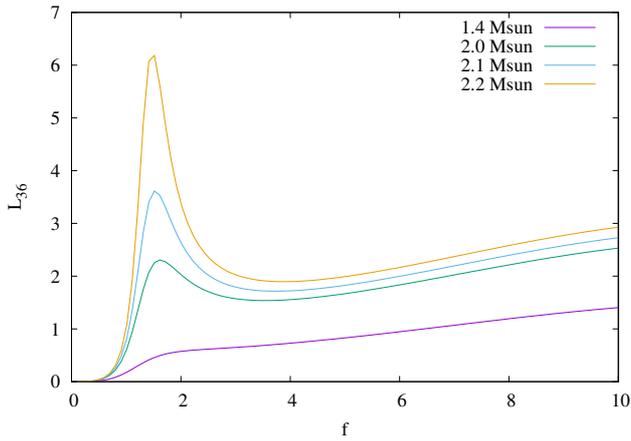}
\caption{X-ray luminosity of XB 1916-053 in units of 10$^{36}$ erg
  s$^{-1}$ versus the thermal bloating factor $f$ of the companion
  star. The four curves correspond to different values of the NS
  mass: purple, green, light blue, and gold colours correspond to a NS
  mass of 1.4, 2, 2.1, and 2.2 M$_{\odot}$, respectively. The peaks in
  the curves are at $f \simeq 1.5$.}
 \label{lx_f}
\end{figure}
We show $L_x$ in unit of $10^{36}$ erg s$^{-1}$ versus $f$ for a NS
mass of  1.4 M$_{\odot}$ (purple colour)  in 
Fig. \ref{lx_f}. Since the observed luminosity for a
NS mass of 1.4 M$_{\odot}$  is larger than the predicted one for each value of 
$f$, also taking  the corresponding error into account, we conclude
that {\bf this specific}
non-conservative mass transfer  scenario fails for a NS mass of 1.4
M$_{\odot}$.      

We  repeat the same procedure for NS masses of 2, 2.1 and
2.2 M$_{\odot}$, finding that the predicted and observed luminosities are
only compatible in the case in which the NS 
 mass is $\ge 2.2$ M$_{\odot}$. In this case ,we find that
$\beta f^{3/2}=0.154$, $\alpha < 0.784$ and  
$$
\bar{x_{L1}}=0.927 -6.02 \times 10^{-2} f + 5.66 \times 10^{-3} f^2 -
2.88   \times 10^{-4} f^3
$$, 
with an accuracy of $2 \times 10^{-3}$. The luminosity for a NS mass of
2.2 M$_{\odot}$ (gold colour) is shown in Fig. \ref{lx_f}. 
\begin{figure}
\centering
\includegraphics[width=6.cm,angle=-90]{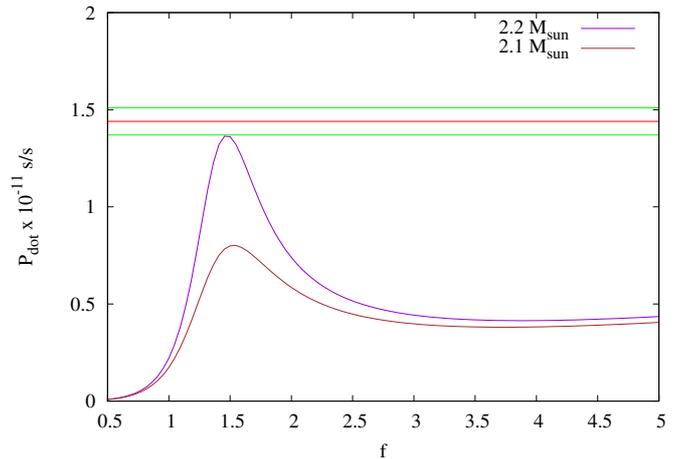}
 \caption{Orbital period derivative of XB 1916-053 in units of
   10$^{-11}$ s s$^{-1}$ versus the thermal bloating factor $f$ of the
   companion star. The brown and purple curves are obtained using a NS
   mass of 2.1 and 2.2 M$_{\odot}$.  The red  and  green lines indicate the best-fit
   value and the values at 68\% confidence level of the orbital period derivative obtained from the LQS
   ephemeris. The purple curve is compatible at 1 $\sigma$ with the
   measured orbital period derivative for $f \simeq 1.5$.  }
 \label{pdot_f}
\end{figure}
Furthermore, we plot the orbital period derivative as function of $f$
for a NS mass of 2.1 M$_{\odot}$ (brown colour) and 2.2 M$_{\odot}$
(purple colour) in Fig.  \ref{pdot_f}.  We note that only for a NS 
 mass of 2.2 M$_{\odot}$ the predicted and measured $\dot{P}$ are
compatible for $f \simeq 1.5$.  We conclude that this
non-conservative mass transfer scenario predicts the observed values
of luminosity and orbital period derivative only for NS
masses larger than 2.2 M$_{\odot}$.  For a NS mass of 2.2
M$_{\odot}$, the companion star has a mass of 0.028 M$_{\odot}$ and
$\beta$ is close to 0.084, which is more than 90\% of the matter, yielded
from the companion star, that leaves the binary system.

In this scenario, we suggest that XB 1916-053 could be considered as a possible
progenitor of the ultra-compact "Black Widow" pulsars with very low-mass companions. \cite{benvenuto_12} proposed that a binary system
with an initial orbital period of 0.8 d, composed  of a 
1.4 M$_{\odot}$ NS and a companion star mass of 2 M$_{\odot}$, evolves in
$\sim$6.5 Gyr forming a binary system that well fits the known orbital
parameters of the black widow millisecond pulsar \object{PSR J1719-1438}. We
note that the same evolutive path fits the orbital parameters of
XB 1916-053 at $\sim 5$ Gyr from the initial time. At  $ 5$ Gyr, 
 the predicted orbital period is  0.035 d, the
predicted companion star mass is 0.03 M$_{\odot}$, the NS
mass is slightly larger than 2.2 M$_{\odot}$  (Benvenuto, private
communication) and the companion star is
helium dominated. These 
values are very similar to those of  XB 1916-053 shown in this
work for a non-conservative mass transfer scenario, although
a discrepancy between our estimation of $\dot{M_2} \sim 4.1 \times
10^{-9}$ M$_{\odot}$ yr$^{-1}$ and the
value suggested by   \cite{benvenuto_12} at 5 Gyr ($\sim 10^{-10}$ M$_{\odot}$ yr$^{-1}$) is present. 
Furthermore, we   note that as the spin period of
PSR J1719-1438 is 5.7 ms \citep[see][and references therein]{bailes}
 the spin period of the NS in XB 1916-053 could also be
extremely short. Indeed, \cite{Galloway2} interpreted the asymptotic frequency of the
coherent burst oscillations in terms of a decoupled surface burning layer and
suggested that  the NS could have a spin period around 3.7 ms.

Nevertheless, we note that our solution for a non-conservative mass
transfer scenario is not supported by a robust physical mechanism to
explain the large quantity of matter ejected from the inner Lagrangian
point.  To date, only two physical mechanisms are known to be able to
eject the transferred matter partially (or totally) . The first mechanism
predicts that when a super-Eddington mass transfer occurs, the X-ray
luminosity has to be at the Eddington limit. Then, the radiation
pressure from the compact object pushes away part of the transferred
matter from the binary system.  This mechanism was recently invoked to
explain the large orbital period derivative measured in the accretion
disc corona (ADC) source \object{X1822-371} by \cite{burderi_10},
\cite{iaria_13}, and \cite{iaria_15}. However, this mechanism cannot be
applied in the case of XB 1916-053 because type-I X-ray bursts are
observed in the light curve of the source (see
e.g. Fig. \ref{lightsuz}), whilst the stable burning sets in at high
accretion rate values that are comparable to the Eddington limit
\citep[see][and references therein]{bild_00}.  Consequently, the
mass transfer rate cannot be super-Eddington and this mechanism cannot
justify a non-conservative mass transfer scenario.  The second
mechanism supposes that the X-ray binary system is a transient source
and during the X-ray quiescence it is ejecting the transferred matter
from the inner Lagrangian point due to the radiation pressure of the
magneto-dipole rotator emission. This mechanism, which we call
radio ejection after \cite{Burderi_01}, was proposed by
\cite{disalvo_08} to explain the large orbital period derivative
measured in \object{SAX J1808.4--3658}. However, this mechanism also fails to
explain our results because  XB 1916-053 is a persistent
X-ray source.

Finally, we discuss the sinusoidal modulation observed in the LQS and
LSe ephemerides.  If we assume a conservative mass transfer scenario,
the predicted orbital period derivative is close to $4 \times 10^{-13}$ s
s$^{-1}$ independent  of the NS mass. 
Then we added a quadratic term to the LSe ephemeris to take
the predicted value into account. 
 We fitted again the delays using the relation
$$
y(t) = a +bt+ct^2+\Delta_{DS}(t),
$$
where the term $c$ is fixed to $5 \times 10^{-7}$ s/d$^2$. The fit
parameters are reported in Tab. \ref{fit_result_LSe_c}.
\begin{table}
\setlength{\tabcolsep}{3pt} 
  \caption{Best-fit  parameters of the delays assuming the presence of the third body in eccentric orbit and
taking a quadratic term $c=5 \times 10^{-7}$ s/d$^2$into account. \label{fit_result_LSe_c}}
\begin{center}
\begin{tabular}{l c c c }          
\hline                                             
\hline  
Parameters &  $P_{mod}$=17$\,$100 d  & $P_{mod}$=18$\,$600 d &  $P_{mod}$=20$\,$100 d \\  
\hline                                             
$a$   (s)    &   $180 \pm 332$     &  $21 \pm 307$ &   $-27 \pm 285$   \\
$b$   ($\times 10^{-3}$ s/d)  &  $2 \pm 20$    &  $2 \pm 19$&  $4 \pm 21$   \\
$A$   (s)      &   $506 \pm 46$    &  $534 \pm 43$&   $562 \pm 43$   \\
$e$             &  $0.26 \pm 0.20$      & $0.28 \pm 0.15$&  $0.32 \pm 0.13$  \\
$\varpi$ (deg)  &  $198 \pm 27$  & $213 \pm 28$&   $219 \pm 27$\\
$t_\phi$ (d)     &   $-3\,594 \pm 1\,129$   & $-3\,036 \pm 1\,131$&   $-2\,825\pm 1\,042$ \\
                                                                    
$\chi^2$(d.o.f.) &51.3(21)   &47.9(21)&45.5(21)\\ 
F-test prob.       &   $3.5 \times 10^{-2}$  &   $1.5 \times 10^{-2}$   &  $0.8 \times 10^{-2}$  \\

\hline                                             

\end{tabular}
\end{center}
{\small \sc Note} \footnotesize---  
The reported errors are at 68\% confidence level. The F-test
probability is estimated with respect to the $\chi^2$ value of the LS
ephemeris (the fourth column of Tab. \ref{fit_result}).
\end{table}
We note that the addition of the quadratic term does not significantly
change the best-fit parameters.

An explanation of the sinusoidal modulation obtained from the LSe
ephemeris could be the presence of a third body gravitationally bound
to the X-ray binary system.  Assuming the existence of a third body of
mass $M_3$, the binary system XB 1916-053 orbits around the new centre
of mass (CM) of the triple system.  The distance of XB 1916-053 from
the new CM is given by $a_x = a_{bin} \sin i = A \; c,$ where $i$ is
the inclination angle of the orbit with respect to the line of sight,
$A$ is the amplitude of the sinusoidal function obtained from the
ephemeris of eq.  \ref{linear_sin_e_eph}, and $c$ is the light speed.
We obtained $a_x = (1.60 \pm 0.13) \times 10^{13}$ cm for
$P_{mod}=18\,600$ d.
We can write the mass function of the triple system as 
$$
\frac{M_3 \sin i}{(M_3+M_{bin})^{2/3}} = \left(\frac{4 \pi^2}{G}\right)^{1/3} \frac{a_x}{P_{mod}^{2/3}},
$$
where $M_3$ is the third body mass, $M_{bin}$ the binary system mass,
and finally, $P_{mod}$ is the orbital period of  XB 1916-053 around
the CM of  the triple system.  
Substituting the values of $M_{bin}$, $P_{mod}$, $a_x$, and assuming an
inclination angle for the source of $70^{\circ}$, we find that $m_3$ is
$ \sim 0.10 $ M$_{\odot}$ and $\sim 0.14 $ M$_{\odot}$ for a NS
mass of 1.4 M$_{\odot}$ and 2.2 M$_{\odot}$, respectively.  We used also
$P_{mod}$ of 17$\;$100 and 20$\;$100 d finding that the values of
$m_3$ are substantially independent of the value of $P_{mod}$.

For a  non-conservative mass transfer scenario, we discuss the
sinusoidal modulation obtained from the LQS ephemeris assuming a
NS mass of 2.2  M$_{\odot}$. In this case we find that  
$a_x = (3.9 \pm 0.5) \times 10^{12}$ cm and  
$m_3 \sim 0.055 $  M$_{\odot}$ for an inclination angle of  $70^{\circ}$.

\section{Conclusions}

We have systematically analysed all the historically
reported X-ray light curves of XB 1916-053, which span 37
years. We find that the previously suggested quadratic ephemeris for
this source no longer fits the dip arrival times.  

We studied the conservative mass transfer scenario of the system, 
finding that the thermal bloating factor of the degenerate companion
star is 3.6 and 3 for a NS mass of 1.4 and 2.2 M$_{\odot}$. In this
scenario, the predicted and observed luminosity are compatible ($\sim
$5-7 $ \times 10^{36}$ erg s$^{-1}$),  although the orbital period derivative is
a factor of 40 smaller than the value of $1.44 \times 10^{-11}$ s
s$^{-1}$ obtained fitting the delays with a quadratic plus a
sinusoidal function (LQS ephemeris). If the conservative mass transfer scenario is
correct, we  conclude that the modulation of the delays
associated with the dip arrivals time are solely due to a sinusoidal
modulation caused by a third body orbiting around the binary system. 
 In this case we estimate
the third body mass is 0.10 and 0.14 M$_{\odot}$  for NS masses of 1.4 and
2.2 M$_{\odot}$, respectively.
The orbital period of the third body around XB 1916-053 is close to 55
yr and the orbit shows an eccentricity $e = 0.28 \pm 0.15$. 

In a non-conservative mass transfer scenario where the mass is ejected
away from the inner Lagrangian point, we find that the observed
luminosity and the orbital period derivative obtained from the LQS
ephemeris are possible only from a NS mass $\ge 2.2$ M$_{\odot}$.  In
this case we obtain that the thermal bloating factor of the degenerate
companion star is $f \simeq 1.5$, the companion star mass is 0.028
M$_{\odot}$, and the fraction of matter yielded by the companion star and
accreting onto the NS is $\beta = 0.084$. In this scenario, the
sinusoidal modulation of the
delays can be explained by the presence of a third body orbiting
around XB 1916-053 with an period of 26 yr. We find that the
third body mass is 0.055 M$_{\odot}$.
Finally, if the non-conservative mass transfer scenario is valid, 
we suggest that XB 1916-053 
and  the ultra-compact black widow system PSR J1719-1438  
could be two different stages of the same evolutive path discussed by  
\cite{benvenuto_12}. If it is true, then  the age of  XB 1916-053 
is close to 5 Gyr, whilst  PSR J1719-1438  is  $\sim$6.5 Gyr old.

\section*{Acknowledgments}
This research has made use of data and/or software provided by the
High Energy Astrophysics Science Archive Research Center (HEASARC),
which is a service of the Astrophysics Science Division at NASA/GSFC
and the High Energy Astrophysics Division of the Smithsonian
Astrophysical Observatory.  We thank the {\it Swift} team duty
scientists and science planners.  
The High-Energy Astrophysics Group of Palermo acknowledges support
from the Fondo Finalizzato alla Ricerca (FFR) 2012/13, project
N. 2012-ATE-0390, founded by the University of Palermo.
This work was partially supported by the Regione Autonoma
della Sardegna through POR-FSE Sardegna 2007-2013, L.R. 7/2007,
Progetti di Ricerca di Base e Orientata, Project N. CRP-60529,
and by the INAF/PRIN 2012-6. We also acknowledge financial
contribution from the agreement ASI-INAF I/037/12/0.
PR acknowledges contract ASI-INAF
I/004/11/0. 
AR gratefully acknowledges Sardinia Regional Government for the
  financial support (P.O.R. Sardegna F.S.E. Operational Programme of
  the Autonomous Region of Sardinia, European Social Fund 2007-2013 -
  Axis IV Human Resources, Objective l.3, Line of Activity l.3.1.)

\bibliographystyle{aa} 
\bibliography{citations}

\begin{thebibliography}{43}
\expandafter\ifx\csname natexlab\endcsname\relax\def\natexlab#1{#1}\fi

\bibitem[{{Bailes} {et~al.}(2011){Bailes}, {Bates}, {Bhalerao}, {Bhat},
  {Burgay}, {Burke-Spolaor}, {D'Amico}, {Johnston}, {Keith}, {Kramer},
  {Kulkarni}, {Levin}, {Lyne}, {Milia}, {Possenti}, {Spitler}, {Stappers}, \&
  {van Straten}}]{bailes}
{Bailes}, M., {Bates}, S.~D., {Bhalerao}, V., {et~al.} 2011, Science, 333, 1717

\bibitem[{{Barret} {et~al.}(1996){Barret}, {Grindlay}, {Strickman}, \&
  {Vedrenne}}]{Barret_96}
{Barret}, D., {Grindlay}, J.~E., {Strickman}, M., \& {Vedrenne}, G. 1996, AAps,
  120, C269

\bibitem[{{Becker} {et~al.}(1977){Becker}, {Smith}, {Swank}, {Boldt}, {Holt},
  {Serlemitsos}, \& {Pravdo}}]{Becker_77}
{Becker}, R.~H., {Smith}, B.~W., {Swank}, J.~H., {et~al.} 1977, ApJ, 216, L101

\bibitem[{{Benvenuto} {et~al.}(2012){Benvenuto}, {De Vito}, \&
  {Horvath}}]{benvenuto_12}
{Benvenuto}, O.~G., {De Vito}, M.~A., \& {Horvath}, J.~E. 2012, \apjl, 753, L33

\bibitem[{{Bildsten}(2000)}]{bild_00}
{Bildsten}, L. 2000, in American Institute of Physics Conference Series, Vol.
  522, American Institute of Physics Conference Series, ed. S.~S. {Holt} \&
  W.~W. {Zhang}, 359--369

\bibitem[{{Boirin} {et~al.}(2004){Boirin}, {Parmar}, {Barret}, {Paltani}, \&
  {Grindlay}}]{Boirin_xmm}
{Boirin}, L., {Parmar}, A.~N., {Barret}, D., {Paltani}, S., \& {Grindlay},
  J.~E. 2004, \aap, 418, 1061

\bibitem[{{Burderi} {et~al.}(2010){Burderi}, {Di Salvo}, {Riggio}, {Papitto},
  {Iaria}, {D'A{\`i}}, \& {Menna}}]{burderi_10}
{Burderi}, L., {Di Salvo}, T., {Riggio}, A., {et~al.} 2010, \aap, 515, A44

\bibitem[{{Burderi} {et~al.}(2001){Burderi}, {Possenti}, {D'Antona}, {Di
  Salvo}, {Burgay}, {Stella}, {Menna}, {Iaria}, {Campana}, \&
  {d'Amico}}]{Burderi_01}
{Burderi}, L., {Possenti}, A., {D'Antona}, F., {et~al.} 2001, \apjl, 560, L71

\bibitem[{{Callanan} {et~al.}(1995){Callanan}, {Grindlay}, \&
  {Cool}}]{Callanan_95}
{Callanan}, P.~J., {Grindlay}, J.~E., \& {Cool}, A.~M. 1995, PASJ, 47, 153

\bibitem[{{Chou} {et~al.}(2001){Chou}, {Grindlay}, \& {Bloser}}]{Chou_01}
{Chou}, Y., {Grindlay}, J.~E., \& {Bloser}, P.~F. 2001, ApJ, 549, 1135

\bibitem[{{Church} {et~al.}(1997){Church}, {Dotani},
  {Ba{\L}uci{\'N}ska-Church}, {Mitsuda}, {Takahashi}, {Inoue}, \&
  {Yoshida}}]{Church_97}
{Church}, M.~J., {Dotani}, T., {Ba{\L}uci{\'N}ska-Church}, M., {et~al.} 1997,
  ApJ, 491, 388

\bibitem[{{Church} {et~al.}(1998){Church}, {Parmar}, {Balucinska-Church},
  {Oosterbroek}, {dal Fiume}, \& {Orlandini}}]{church_98}
{Church}, M.~J., {Parmar}, A.~N., {Balucinska-Church}, M., {et~al.} 1998, \aap,
  338, 556

\bibitem[{{Courvoisier} {et~al.}(2003){Courvoisier}, {Walter}, {Beckmann},
  {Dean}, {Dubath}, {Hudec}, {Kretschmar}, {Mereghetti}, {Montmerle},
  {Mowlavi}, {Paltani}, {Preite Martinez}, {Produit}, {Staubert}, {Strong},
  {Swings}, {Westergaard}, {White}, {Winkler}, \& {Zdziarski}}]{cour_03}
{Courvoisier}, T.~J.-L., {Walter}, R., {Beckmann}, V., {et~al.} 2003, \aap,
  411, L53

\bibitem[{{Di Salvo} {et~al.}(2008){Di Salvo}, {Burderi}, {Riggio}, {Papitto},
  \& {Menna}}]{disalvo_08}
{Di Salvo}, T., {Burderi}, L., {Riggio}, A., {Papitto}, A., \& {Menna}, M.~T.
  2008, \mnras, 389, 1851

\bibitem[{{Galloway} {et~al.}(2001){Galloway}, {Chakrabarty}, {Muno}, \&
  {Savov}}]{Galloway2}
{Galloway}, D.~K., {Chakrabarty}, D., {Muno}, M.~P., \& {Savov}, P. 2001, ApJL,
  549, L85

\bibitem[{{Galloway} {et~al.}(2008){Galloway}, {Muno}, {Hartman}, {Psaltis}, \&
  {Chakrabarty}}]{Galloway_08}
{Galloway}, D.~K., {Muno}, M.~P., {Hartman}, J.~M., {Psaltis}, D., \&
  {Chakrabarty}, D. 2008, \apjs, 179, 360

\bibitem[{{Grindlay}(1989)}]{Grindlay_89}
{Grindlay}, J.~E. 1989, in ESA Special Publication, Vol. 296, Two Topics in
  X-Ray Astronomy, Volume 1: X Ray Binaries. Volume 2: AGN and the X Ray
  Background, ed. J.~{Hunt} \& B.~{Battrick}, 121--126

\bibitem[{{Grindlay}(1992)}]{Grindlay_92}
{Grindlay}, J.~E. 1992, in Frontiers Science Series, ed. Y.~{Tanaka} \&
  K.~{Koyama}, 69

\bibitem[{{Grindlay} {et~al.}(1988){Grindlay}, {Bailyn}, {Cohn}, {Lugger},
  {Thorstensen}, \& {Wegner}}]{Grindlay_88}
{Grindlay}, J.~E., {Bailyn}, C.~D., {Cohn}, H., {et~al.} 1988, ApJL, 334, L25

\bibitem[{{Grindlay} {et~al.}(1987){Grindlay}, {Cohn}, \&
  {Schmidtke}}]{Grindlay_87}
{Grindlay}, J.~E., {Cohn}, H., \& {Schmidtke}, P. 1987, IauCirc, 4393, 1

\bibitem[{{Hu} {et~al.}(2008){Hu}, {Chou}, \& {Chung}}]{Hu_08}
{Hu}, C.-P., {Chou}, Y., \& {Chung}, Y.-Y. 2008, \apj, 680, 1405

\bibitem[{{Iaria} {et~al.}(2013){Iaria}, {Di Salvo}, {D'A{\`i}}, {Burderi},
  {Mineo}, {Riggio}, {Papitto}, \& {Robba}}]{iaria_13}
{Iaria}, R., {Di Salvo}, T., {D'A{\`i}}, A., {et~al.} 2013, \aap, 549, A33

\bibitem[{{Iaria} {et~al.}(2006){Iaria}, {Di Salvo}, {Lavagetto}, {Robba}, \&
  {Burderi}}]{Iaria_06}
{Iaria}, R., {Di Salvo}, T., {Lavagetto}, G., {Robba}, N.~R., \& {Burderi}, L.
  2006, \apj, 647, 1341

\bibitem[{{Iaria} {et~al.}(2015){Iaria}, {Di Salvo}, {Matranga}, {Galiano},
  {D'A{\'{\i}}}, {Riggio}, {Burderi}, {Sanna}, {Ferrigno}, {Del Santo},
  {Pintore}, \& {Robba}}]{iaria_15}
{Iaria}, R., {Di Salvo}, T., {Matranga}, M., {et~al.} 2015, \aap, 577, A63

\bibitem[{{Lund} {et~al.}(2003){Lund}, {Budtz-J{\o}rgensen}, {Westergaard},
  {Brandt}, {Rasmussen}, {Hornstrup}, {Oxborrow}, {Chenevez}, {Jensen},
  {Laursen}, {Andersen}, {Mogensen}, {Rasmussen}, {Om{\o}}, {Pedersen},
  {Polny}, {Andersson}, {Andersson}, {K{\"a}m{\"a}r{\"a}inen}, {Vilhu},
  {Huovelin}, {Maisala}, {Morawski}, {Juchnikowski}, {Costa}, {Feroci},
  {Rubini}, {Rapisarda}, {Morelli}, {Carassiti}, {Frontera}, {Pelliciari},
  {Loffredo}, {Mart{\'{\i}}nez N{\'u}{\~n}ez}, {Reglero}, {Velasco}, {Larsson},
  {Svensson}, {Zdziarski}, {Castro-Tirado}, {Attina}, {Goria}, {Giulianelli},
  {Cordero}, {Rezazad}, {Schmidt}, {Carli}, {Gomez}, {Jensen}, {Sarri},
  {Tiemon}, {Orr}, {Much}, {Kretschmar}, \& {Schnopper}}]{lund_03}
{Lund}, N., {Budtz-J{\o}rgensen}, C., {Westergaard}, N.~J., {et~al.} 2003,
  \aap, 411, L231

\bibitem[{{Nelemans} {et~al.}(2006){Nelemans}, {Jonker}, \&
  {Steeghs}}]{nelemans}
{Nelemans}, G., {Jonker}, P.~G., \& {Steeghs}, D. 2006, \mnras, 370, 255

\bibitem[{{Paczynski} \& {Sienkiewicz}(1981)}]{Pac}
{Paczynski}, B. \& {Sienkiewicz}, R. 1981, ApJL, 248, L27

\bibitem[{{Priedhorsky} \& {Terrell}(1984)}]{Priedhorsky}
{Priedhorsky}, W.~C. \& {Terrell}, J. 1984, ApJ, 280, 661

\bibitem[{{Rappaport} {et~al.}(1982){Rappaport}, {Joss}, \&
  {Webbink}}]{rappa_82}
{Rappaport}, S., {Joss}, P.~C., \& {Webbink}, R.~F. 1982, \apj, 254, 616

\bibitem[{{Rappaport} {et~al.}(1987){Rappaport}, {Ma}, {Joss}, \&
  {Nelson}}]{rappa_87}
{Rappaport}, S., {Ma}, C.~P., {Joss}, P.~C., \& {Nelson}, L.~A. 1987, \apj,
  322, 842

\bibitem[{{Retter} {et~al.}(2002){Retter}, {Chou}, {Bedding}, \&
  {Naylor}}]{Retter_02}
{Retter}, A., {Chou}, Y., {Bedding}, T.~R., \& {Naylor}, T. 2002, MNRAS, 330,
  L37

\bibitem[{Shapiro \& Teukolsky(1983)}]{Shapiro_83}
Shapiro, S.~L. \& Teukolsky, S.~A. 1983, Black Holes, White Dwarfs, and Neutron
  Stars: the Physics of Compact Objects (John Wiley)

\bibitem[{{Smale} {et~al.}(1988){Smale}, {Mason}, {White}, \&
  {Gottwald}}]{Smale_88}
{Smale}, A.~P., {Mason}, K.~O., {White}, N.~E., \& {Gottwald}, M. 1988, MNRAS,
  232, 647

\bibitem[{{Smale} {et~al.}(1989){Smale}, {Mason}, {Williams}, \&
  {Watson}}]{Smale_89}
{Smale}, A.~P., {Mason}, K.~O., {Williams}, O.~R., \& {Watson}, M.~G. 1989,
  PASJ, 41, 607

\bibitem[{{Swank} {et~al.}(1984){Swank}, {Taam}, \& {White}}]{Swank_84}
{Swank}, J.~H., {Taam}, R.~E., \& {White}, N.~E. 1984, ApJ, 277, 274

\bibitem[{{van der Klis} \& {Bonnet-Bidaud}(1984)}]{vdk_84}
{van der Klis}, M. \& {Bonnet-Bidaud}, J.~M. 1984, \aap, 135, 155

\bibitem[{{Walter} {et~al.}(1982){Walter}, {Mason}, {Clarke}, {Halpern},
  {Grindlay}, {Bowyer}, \& {Henry}}]{Walter_82}
{Walter}, F.~M., {Mason}, K.~O., {Clarke}, J.~T., {et~al.} 1982, ApJ, 253, L67

\bibitem[{{White}(1989)}]{White_89}
{White}, N.~E. 1989, Aapr, 1, 85

\bibitem[{{White} \& {Swank}(1982)}]{White_82}
{White}, N.~E. \& {Swank}, J.~H. 1982, ApJ, 253, L61

\bibitem[{{Winkler} {et~al.}(2003){Winkler}, {Courvoisier}, {Di Cocco},
  {Gehrels}, {Gim{\'e}nez}, {Grebenev}, {Hermsen}, {Mas-Hesse}, {Lebrun},
  {Lund}, {Palumbo}, {Paul}, {Roques}, {Schnopper}, {Sch{\"o}nfelder},
  {Sunyaev}, {Teegarden}, {Ubertini}, {Vedrenne}, \& {Dean}}]{winkler_03}
{Winkler}, C., {Courvoisier}, T.~J.-L., {Di Cocco}, G., {et~al.} 2003, \aap,
  411, L1

\bibitem[{{Yoshida}(1993)}]{Yoshida_PhD_93}
{Yoshida}, K. 1993, PhD thesis, Thesis, Tokyo University, (1993)

\bibitem[{{Yoshida} {et~al.}(1995){Yoshida}, {Inoue}, {Mitsuda}, {Dotani}, \&
  {Makino}}]{Yoshida_95}
{Yoshida}, K., {Inoue}, H., {Mitsuda}, K., {Dotani}, T., \& {Makino}, F. 1995,
  PASJ, 47, 141

\bibitem[{{Zhang} {et~al.}(2014){Zhang}, {Makishima}, {Sakurai}, {Sasano}, \&
  {Ono}}]{zhang_2014}
{Zhang}, Z., {Makishima}, K., {Sakurai}, S., {Sasano}, M., \& {Ono}, K. 2014,
  \pasj, 66, 120

\end{thebibliography}
\end{document}